\documentclass[%
reprint,
 amsmath,amssymb,
 aps,
]{revtex4-2}

\usepackage{graphicx}%
\usepackage{dcolumn}%
\usepackage{bm}%
\usepackage[hidelinks]{hyperref}%
\usepackage{xcolor}
\usepackage{booktabs}

\begin{document}

\preprint{APS/123-QED}

\title{Reconstruction of Three-dimensional Scroll {Waves in Excitable Media \\from Two-Dimensional Observations} using Deep Neural Networks}

\author{Jan Lebert}
 \affiliation{Cardiovascular Research Institute, University of California, San Francisco, USA}%

\author{Meenakshi Mittal}
 \affiliation{Cardiovascular Research Institute, University of California, San Francisco, USA}

\author{Jan Christoph}
 \homepage{https://cardiacvision.ucsf.edu}
 \email{jan.christoph@ucsf.edu}
\affiliation{Cardiovascular Research Institute, University of California, San Francisco, USA}

\begin{abstract}

Scroll wave chaos is thought to underlie life-threatening ventricular fibrillation. 
However, currently there is no direct way to measure action potential wave patterns transmurally throughout the thick ventricular heart muscle.
Consequently, direct observations of three-dimensional electrical scroll waves remains elusive. 
{Here, we study whether it is possible to reconstruct simulated scroll waves and scroll wave chaos using deep learning.} 
We trained encoding-decoding convolutional neural networks to predict three-dimensional {scroll wave dynamics} inside bulk-shaped excitable media from two-dimensional observations of the wave dynamics on the bulk's surface.
We tested whether observations from one or two opposing surfaces would be sufficient, and whether {transparency or} measurements of surface deformations enhances the reconstruction. 
Further, we {evaluated the approach's robustness against noise} and tested the feasibility of predicting the bulk's thickness.
{We distinguished isotropic and anisotropic, as well as opaque and transparent excitable media as models for cardiac tissue and the Belousov-Zhabotinsky chemical reaction, respectively.} 
{While we demonstrate that it is possible to reconstruct three-dimensional scroll wave dynamics, we also show that it is challenging to reconstruct complicated scroll wave chaos and that prediction outcomes depend on various factors such as transparency, anisotropy and ultimately the thickness of the medium compared to the size of the scroll waves.
In particular,} we found that anisotropy provides crucial information for neural networks to decode depth, which facilitates the reconstructions.
In the future, deep neural networks could be used to visualize intramural action potential wave patterns from epi- or endocardial measurements.

\begin{description}
\item[Keywords]
Scroll waves, excitable media, action potential waves, ventricular fibrillation, deep learning
\end{description}

\end{abstract}

\maketitle

\section{Introduction}
Scroll wave {dynamics} occur in excitable reaction-diffusion systems, termed 'excitable media'. 
They are conjectured to underlie life-threatening heart rhythm disorders, such as ventricular fibrillation. 
In the heart, nonlinear waves of electrical excitation propagate through the cardiac muscle and initiate its contractions.
The electrical waves are conjectured to degenerate into electrical scroll wave chaos via a cascade of wavebreaks during the onset of ventricular fibrillation.
However, direct evidence for the existence of scroll waves in the heart is lacking. 
While the dynamics of scroll waves have been studied extensively in computer simulations \cite{Fenton1998, Clayton2008, Pathmanathan2015}, the direct visualization of scroll waves throughout the depths of the heart muscle remains a challenge. %

Spiral wave-like action potential waves can be imaged on the heart surface during ventricular tachycardia or fibrillation using voltage-sensitive optical mapping \cite{Davidenko1992, Pertsov1993, Winfree1994, Christoph2018, Uzelac2022}, %
and the surface observations are in agreement with simulated three-dimensional scroll wave dynamics \cite{Pathmanathan2015}.
Otherwise, only few and indirect experimental evidence of scroll waves in the heart exists.
Voltage-sensitive transillumination imaging was used to measure projections of scroll waves on the surface of the isolated right ventricle of porcine and sheep hearts \cite{Baxter2001, Bernus2007, Mitrea2009}.
The right ventricles are thinner than the left and can therefore be penetrated ($\sim 0.5\,\text{cm}$) by near-infrared light, {making them semi-transparent}.
Consequently, it was possible to locate focal wave sources inside the volume of the right ventricle using transillumination imaging \cite{Khait2006,Caldwell2015}.
More recently, it was shown that ultrasound imaging can reveal mechanical vortices in the ventricles of whole isolated porcine hearts, which co-exist with electrical vortices on the epicardial surface, suggesting that the heart's mechanical dynamics reflect electrical scroll wave dynamics \cite{Christoph2018, Molavi2022}.
However, it remains difficult to extrapolate the measured projections or surface observations of the electrical dynamics into the depths of the cardiac muscle, and eventually correlate them with mechanical measurements.
Fully three-dimensional reconstructions of scroll waves were obtained using optical tomography in the transparent Belousov-Zhabotinsky chemical reaction \cite{Welsh1983,Bansagi2008,Daehmlow2013, Bruns2017}, {which is an excitable medium that shares very similar wave dynamics with cardiac tissue.}
In contrast, three-dimensional action potential waves have been directly measured in small rat and zebrafish hearts using laminar optical tomography \cite{Hillman2007} or light-sheet microscopy \cite{Sacconi2022}.
However, attempts to obtain three-dimensional visualizations of scroll waves inside the optically dense cardiac muscle of large mammalian hearts have not yet attained a similar quality, and better measurement and reconstruction techniques are needed.

Multiple numerical approaches for the reconstruction of scroll waves from surface observations have previously been proposed:
Berg et al.\ \cite{Berg2011} attempted to recover simulated scroll wave chaos from single-surface observations using a synchronization-based data-assimilation approach. 
However, while the approach was successful at recovering scroll wave chaos from sparse measurements within the medium, see also \cite{Lebert2019}, it was not suited to extrapolate scroll wave dynamics into the three-dimensional bulk-shaped medium from surface observations.
Hoffman et al.\ \cite{Hoffman2016, Hoffman2020} analyzed dual-surface observations (comparable to measuring both the epi- and {endocardium}) using a different data-assimilation approach, the local ensemble transform Kalman filter \cite{Hunt2007}, to successfully reconstruct simulated scroll waves.
The question remains if it is possible to reconstruct truly complex three-dimensional scroll wave chaos from single- or dual-surface observations.
More recently, neural networks were used to predict cardiac dynamics from sparse or partial observations with promising results \cite{Lebert2021,Herzog2021,HerreroMartin2022}.
However, the task of predicting scroll wave dynamics from surface observations using neural networks has not yet been established.

\begin{figure}[htb]
  \centering
  \includegraphics[width=0.48\textwidth]{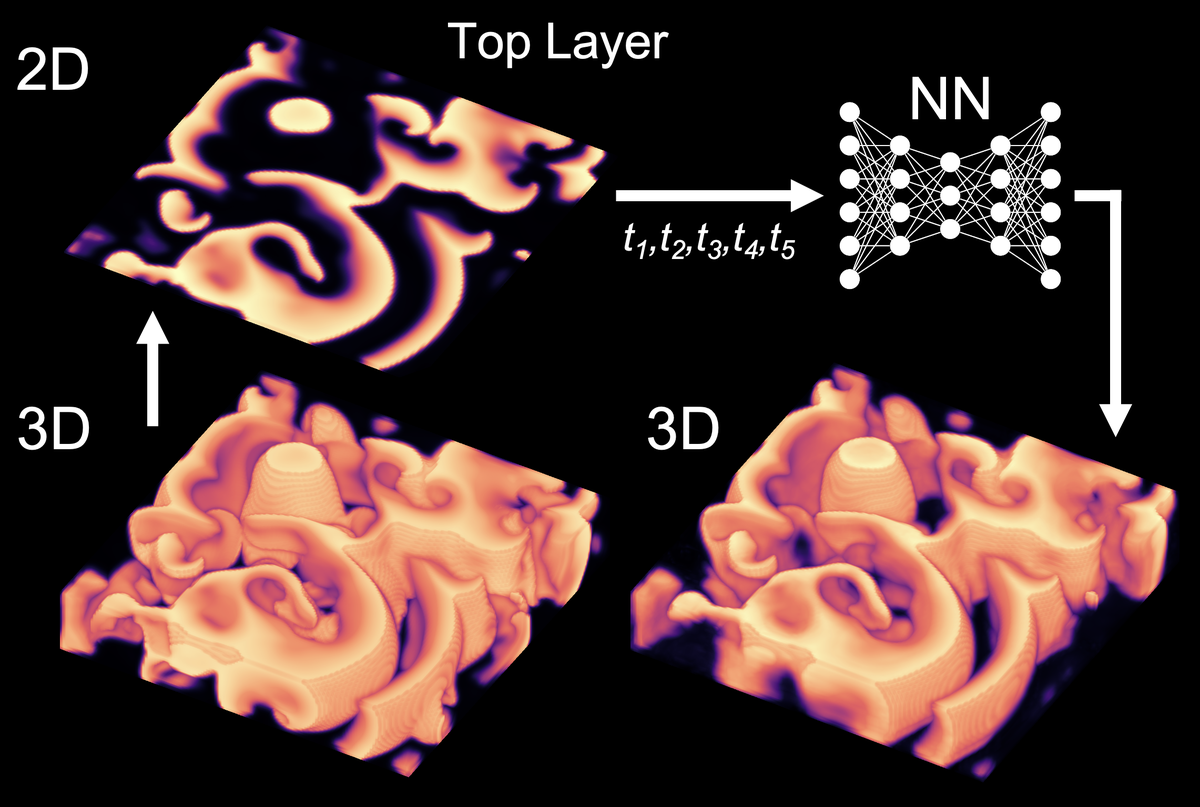}
  \caption{
Deep learning-based reconstruction of scroll wave chaos inside a three-dimensional volume from partial observations of the dynamics on its surface. 
Scroll wave chaos is a model for the electrophysiological dynamics underlying ventricular fibrillation. 
Computer simulations were performed in isotropic and anisotropic bulk-shaped excitable media. 
A neural network (NN) predicts scroll wave dynamics underneath the bulk's surface from a short temporal sequence ($t_1, \ldots, t_5$) of two-dimensional observations (here shown for top layer).
  }
  \label{fig:Figure01}
\end{figure}

Here, we provide a numerical proof-of-principle that deep encoding-decoding convolutional neural networks, under certain conditions, can be used to reconstruct three-dimensional {scroll wave dynamics, including complicated scroll wave chaos,} from two-dimensional observations of the dynamics, {see Fig.~\ref{fig:Figure01}}.
We show that scroll waves can be recovered fully when the size of the waves is in the order of the thickness of the medium, or when analyzing projections of dynamics with more and smaller waves in transparent anisotropic excitable media.
We tested several deep convolutional neural network architectures and analyzed their reconstruction performance depending on opacity, thickness and anisotropy of simulated excitable media.

\section{Methods}
We performed simulations of three-dimensional electrical and electromechanical scroll wave {dynamics} in bulk-shaped isotropic and anisotropic (elastic) excitable media, respectively, 
and used neural networks to predict the three-dimensional wave patterns from a short sequence of two-dimensional observations of the dynamics on the bulk's surface.
{We distinguished `laminar' scroll wave dynamics consisting of 1-3 meandering scroll waves and `turbulent' scroll wave chaos.}

\begin{figure*}[htb]
  \centering
  \includegraphics[width=0.95\textwidth]{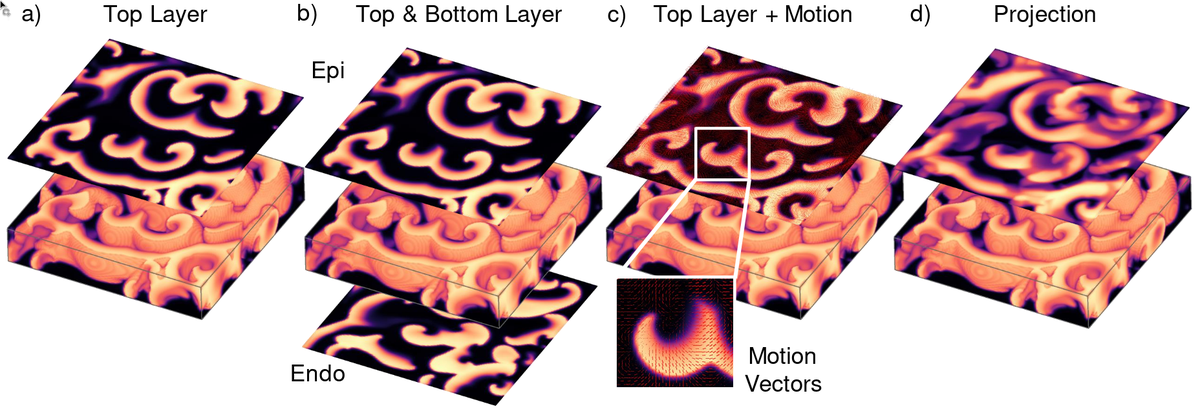}
  \caption{
Surface observations and projections of three-dimensional scroll wave chaos {(`turbulent' parameter regime)} in a bulk medium.
a) Observation of the top surface of the bulk (layer 1) in single-surface mode. 
b) Observation of the top and bottom surfaces of the bulk (layers 1 and 24) in dual-surface mode. 
c) Observation of the electrical activity and mechanical motion on the top surface of the bulk (layer 1) in single-surface mode (red: motion vectors). 
In a-c), the medium is opaque and does not allow observations of the dynamics inside the medium below the top layer.
d) Observation of the projection (transillumination) of the three-dimensional dynamics along the depth of the bulk. The projection is calculated by summing the values in all 24 layers along the $z$-axis for a specific $(x,y)$-coordinate and dividing the sum by the number of layers.
All layers $1-24$ are cross-sections in the $x-y$ plane.
  }
  \label{fig:Figure02}
\end{figure*}

\subsection{Scroll Wave {Dynamics} in Elastic Excitable Media}
\label{sec:methods:simulations}

\begin{figure}[htb]
  \centering
  \includegraphics[width=0.49\textwidth]{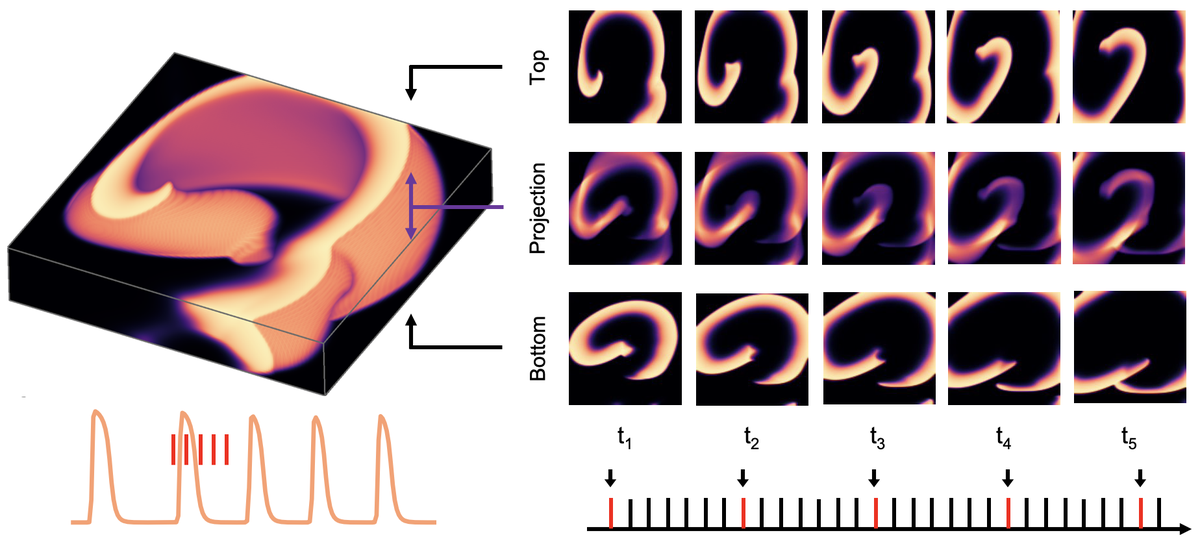}
  \caption{
{Observations of three-dimensional scroll wave (`laminar' parameter set) on the top and bottom surfaces of an opaque medium, and in the projection of the full dynamics in a transparent medium in the bulk's $z$-direction (depth). In each case, the neural network analyzes a short sequence of $5$ snapshots, which are sampled at discrete times (red) over the period of the scroll wave from the simulation data, see section \ref{sec:methods:deeplearning} for details. In the simulations, one rotational period corresponds to about $500$ simulation time steps.}
  }
  \label{fig:ScrollWaveMethod}
\end{figure}

We simulated electrical scroll wave {dynamics} in bulk-shaped excitable media of size $128 {\times} 128 {\times} d_z$ voxels with varying thicknesses or depths {$d_z \in \{ 8, \ldots, 40\}$}.
We used the phenomenological Aliev-Panfilov model \cite{AlievPanfilov1996} to simulate nonlinear waves of electrical excitation: 
\begin{eqnarray} 
\label{eq:modelu}
\frac{\partial u}{\partial t} & = & \nabla \cdot (\bm{D} \nabla u) - k u (u-a) (u-1) - u r \\
\label{eq:modelr}
\frac{\partial r}{\partial t} & = & \epsilon(u,r)(k u(a+1-u)-r)
\end{eqnarray}
The dynamic variables $u$ and $r$ represent the local electrical excitation (voltage) and refractory state, respectively, and are dimensionless, normalized units. 
Together with the term $\epsilon(u,r) = \epsilon_0 + \mu_1 r / (u+\mu_2)$, the partial differential equations describe the local excitable kinetics and diffusive dynamics. 
The parameters $k$, $a$, $\epsilon_0$, $\mu_1$ and $\mu_2${, listed in Table~\ref{tab:parameters},} influence the properties of the excitation waves.
The partial differential equations were integrated using the forward Euler method in a finite differences numerical integration scheme {and used Neumann (zero-flux) boundary conditions}.
We simulated both isotropic
\begin{eqnarray}
  \bm{D} &=& D_{\text{iso}} \bm{I}
\end{eqnarray}
and anisotropic excitable media with locally varying fiber direction with diffusion coefficients for the parallel $D_{\parallel} ${fiber direction and perpendicular $D_{\perp 1}, D_{\perp 2}$ to it \cite{Fenton1998}:}
\begin{eqnarray}
  \bm{D} &=& \begin{pmatrix}
    D_{11} & D_{12} & 0\\
    D_{21} & D_{22} & 0\\
    0 & 0 & D_{33}
    \end{pmatrix} \\\nonumber
    D_{11} &=& D_{\parallel} \cos^2 (\theta(z)) + D_{\perp 1} \sin^2 (\theta(z))\\\nonumber
    D_{22} &=& D_{\parallel} \sin^2 (\theta(z)) + D_{\perp 1} \cos^2 (\theta(z))\\\nonumber
    D_{12} &=& D_{21} = (D_{\parallel} - D_{\perp 1}) \cos (\theta(z)) \sin(\theta(z))\\\nonumber
    D_{33} &=& D_{\perp 2}
\end{eqnarray}
Here, the fiber organization represents ventricular muscle tissue with muscle fibers aligned in sheets in the $x$--$y$~plane and the sheet-fiber orientation rotating throughout the thickness of the bulk.
{$D_{\perp 1}$ is the diffusivity perpendicular to the fiber axis in the $x$--$y$~plane and $D_{\perp 2}$ transmurally, for simplicity we set $D_{\perp} = D_{\perp 1} = D_{\perp 2}$.}
We used a varying fiber angle $\theta(z)$ ranging from $0^{\circ}$ to $90^{\circ}$ between the top and bottom layer of the bulk for the simulation depth $d_z = 24$ voxels:
\begin{eqnarray}
  \theta(z) = z \cdot \Delta \theta
\end{eqnarray}
For the other depths, $d_z \in \{ 8, 12, 16, 20, 28, 32, 40\}$, we used the same $\Delta \theta$ as in the $d_z = 24$ case.
The ratio between the parallel $D_{\parallel}$ and perpendicular $D_{\perp}$ diffusion coefficients was set to $4:1$.
We chose $D_{\perp} = D_{\text{iso}} = 0.05$, {and adapted $\Delta t$ for the Euler integration such that $500$ simulation time steps correspond to about $0.5 - 1.0$ scroll wave rotations. 
The simulation time steps required for one scroll wave rotation fluctuates and depends on the parameter values as well as an/-isotropy.
To save disk space, we stored only every $80th$ simulation time step as one `snapshot', such that $5$ snapshots covered about a half to one scroll wave rotation, see also Fig.~\ref{fig:ScrollWaveMethod}.}

In addition to the purely electric simulations, we also simulated electromechanical scroll wave {dynamics} in deforming excitable media, as described in \cite{Lebert2019}.
{
In short, we coupled a three-dimensional mass-spring damper system with hexahedral cells and tunable fiber anisotropy \cite{Bourguignon2000} to the electric simulation.
Each cell in the mechanical part of the simuation corresponded to one cell or voxel in the electrical part of the simulation. 
Active tension generation in each cell was modelled using an active stress variable $T_a$ that is directly dependent on the excitation variable $u$, as described in \cite{NashPanfilov2004}:
\begin{eqnarray} 
\label{eq:modelT}
\frac{\partial T_a}{\partial t} &=& \epsilon (u)\cdot(k_T u - T_a)\\
\epsilon (u) &=& \begin{cases}
  10  & \text{if } u < 0.05 \\
  1 & \text{if } u \geq 0.05
\end{cases}\nonumber
\end{eqnarray}
The axis along which each cell exerts active contractile force could be pointed into an arbitrary direction, and, throughout the bulk, the axis alignment matched the rotating orthotropic fiber alignment already defined in the electrical part of the simulation.
The mechanical parameter $k_T$ and other parameters shown in Table~\ref{tab:parameters} influence the magnitude of contraction and the properties of the elasticity of the mass-spring damper system.
The elastic medium's boundaries were non-rigid and confined by elastic springs acting on the medium's boundary, see \cite{Lebert2019} for details.
In general, the electromechanical simulation produces three-dimensional deformation patterns that are highly correlated with the electrical scroll wave chaos.}

\begin{table}[htb]
  \begin{tabular}{@{}ccc@{}}\toprule
    Parameter    & `Laminar' Set  & `Turbulent' Set \\\midrule
    $k$          & $8$     & $8$     \\
    $a$          & $0.05$  & $0.05$  \\
    $\epsilon_0$ & $0.002$ & $0.002$ \\
    $\mu_1$      & $0.8$   & $0.2$   \\
    $\mu_2$      & $0.3$   & $0.3$   \\\midrule
    $k_T$        & ---     & $3$     \\
    $k_{ij}$     & ---     & $5$     \\
    $k_j$        & ---     & $0.5$   \\
    $k_f$        & ---     & $4$     \\ 
    $c_f$        & ---     & $10$    \\\bottomrule
  \end{tabular}
  \caption{{
    Electrical (top) and mechanical (bottom) parameters used to simulate two different regimes of scroll wave dynamics: a `laminar' regime, see Figs.~\ref{fig:ScrollWave}, \ref{fig:LargeScrollWave}, and a fully `turbulent' scroll wave chaos regime, see Figs~\ref{fig:Figure03}(right), \ref{fig:Figure04} and \ref{fig:scrollwavechaoslayers}.
    Electromechanical simulations were only performed with the `turbulent' parameter set.
    }}
  \label{tab:parameters}
\end{table}

\begin{figure*}[htb]
  \centering
  \includegraphics[width=0.98\textwidth]{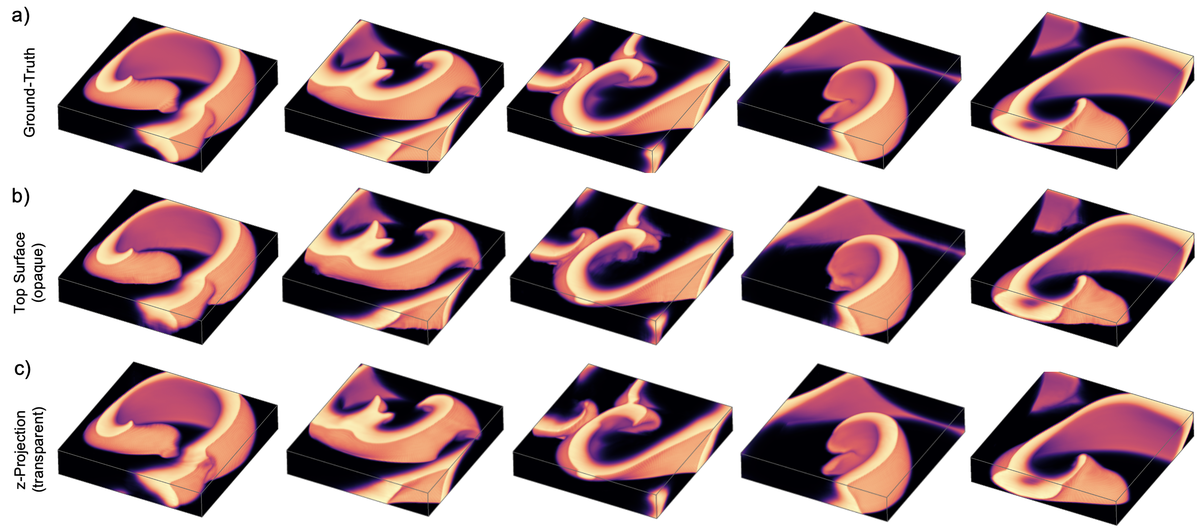}
  \caption{
{Predictions of three-dimensional `laminar' scroll wave dynamics from two-dimensional observations using deep convolutional encoding-decoding neural network (U-Net) in an anisotropic excitable medium ($128 \times 128 \times 24$ voxels).
a) Ground-truth scroll wave dynamics (5 random representative snapshots). The simulations exhibit scroll waves with meandering and curved vortex cores, see also Fig.~\ref{fig:LargeScrollWave}d), wavebreak, and dissociation between the top and bottom surface dynamics, see Fig.~\ref{fig:Figure03}.
b) Predictions from two-dimensional wave pattern visible only on the top surface of the bulk when the medium is completely opaque. The reconstruction accuracy decreases slightly with increasing depth (wave pattern becomes fuzzy towards the bottom).
c) Predictions from two-dimensional projection of the whole three-dimensional dynamics along the $z$-axis in a transparent medium. The prediction accuracy is slightly better than in b), particularly towards the bottom layers of the bulk.
Overall, the predictions and the ground-truth are visually difficult to distinguish from each other. 
The data was not seen by the network during training.
See corresponding Supplementary Movie 1 as well as Supplementary Movies 2, 3 and 4 for a comparison with more complicated scroll wave dynamics.}
  }
  \label{fig:ScrollWave}
\end{figure*}

\begin{figure}[htb]
  \centering
  \includegraphics[width=0.48\textwidth]{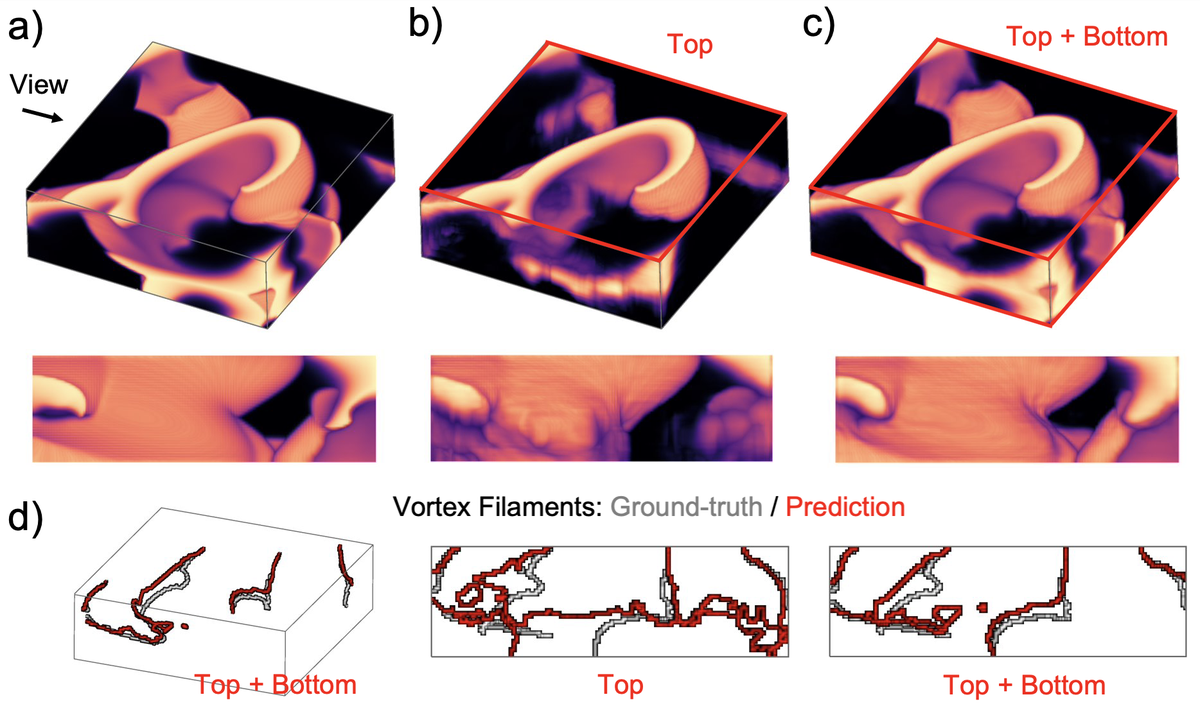}
  \caption{
{Predictions of three-dimensional scroll wave dynamics (`laminar' parameter set) and their vortex filaments from either single- or dual-surface observations in a thick, opaque, anisotropic excitable medium ($128 \times 128 \times 40$ voxels), see also corresponding Supplementary Movie 2.
a) Ground-truth scroll wave dynamics.
b) Prediction from the top surface only. 
c) Prediction from both the top and bottom surfaces. In dual-surface mode, the network is able to recover the dynamics sufficiently well.
Arrow indicates direction of cross-sectional view.
d) Ground-truth vortex filaments (gray) and vortex filaments calculated from predicted scroll wave dynamics (red).
In single-surface mode, predictions become unreliable (wave pattern becomes fuzzy / vortex filaments do not match) towards the bottom of the bulk.
Reconstructions performed with U-Net.}
}
  \label{fig:LargeScrollWave}
\end{figure}

{
We simulated two different regimes of scroll wave dynamics: 1) a `laminar' regime with $1-3$ meandering scroll waves with wavebreaks as shown in Figs.~\ref{fig:ScrollWave} and \ref{fig:LargeScrollWave} and 2) a fully `turbulent' scroll wave chaos regime as shown in Figs.~\ref{fig:Figure04}-\ref{fig:scrollwavechaoslayers}, see Table~\ref{tab:parameters} for the respective parameter values.
For both parameter regimes, we performed $125$ isotropic and anisotropic simulations of electrical scroll wave chaos for different bulk depths -- $d_z \in \{16, 24, 32, 40\}$ voxels for the `laminar' regime and $d_z \in \{ 8, 12, 16, 20, 24, 28, 32 \}$ for the `turbulent' regime. $100$ simulations were used during generation of the training dataset and $25$ simulations were exclusively used for evaluation.
Furthermore, we performed $125$ simulations of electromechanical scroll wave chaos for a thickness of $d_z = 24$, with the same split between training and evaluated dataset.
The electrical and mechanical parameters were identical in each simulation.
However, the initial conditions $u_{t=0}(x, y, z), r_{t=0}(x, y, z)$ were randomized and therefore different in each simulation.
We used cross-field stimulation to set $u_{t=0}, r_{t=0}$ such that two scroll waves are induced at random positions $(x_1, y_1), (x_2, y_2)$.
Additionally we added a small amount of Gaussian noise (standard deviation $\sigma = 0.1$) to the initial conditions $u_{t=0}, r_{t=0}$.
With both parameter sets, the dynamics quickly diverged.
We discarded the first $75$ snapshots of each simulation (approximately $15$ scroll wave rotation periods) and used the remaining $500$ snapshots (approximately $100$ scroll wave rotation periods) for generating the training or evaluation data, respectively.
If the excitation in a simulation decayed ($u_t \approx 0~\forall (x,y,z)$), we restarted the simulation with new initial conditions $u_{t=0}, r_{t=0}$.
The training dataset consists of $20{,}000$ randomly selected samples from the $100$ training simulations, and the evaluation dataset uses $5{,}000$ randomly chosen samples from the $25$ evaluation simulations.
Consequently, training and evaluation datasets were completely separate datasets.
}
The numerical simulation was implemented in C++, the source code for the simulations is available in \cite{Lebert2019}.

\subsection{Deep Learning-based Reconstruction of Scroll Wave {Dynamics}}\label{sec:methods:deeplearning}

\begin{figure}[htb]
  \centering
  \includegraphics[width=0.42\textwidth]{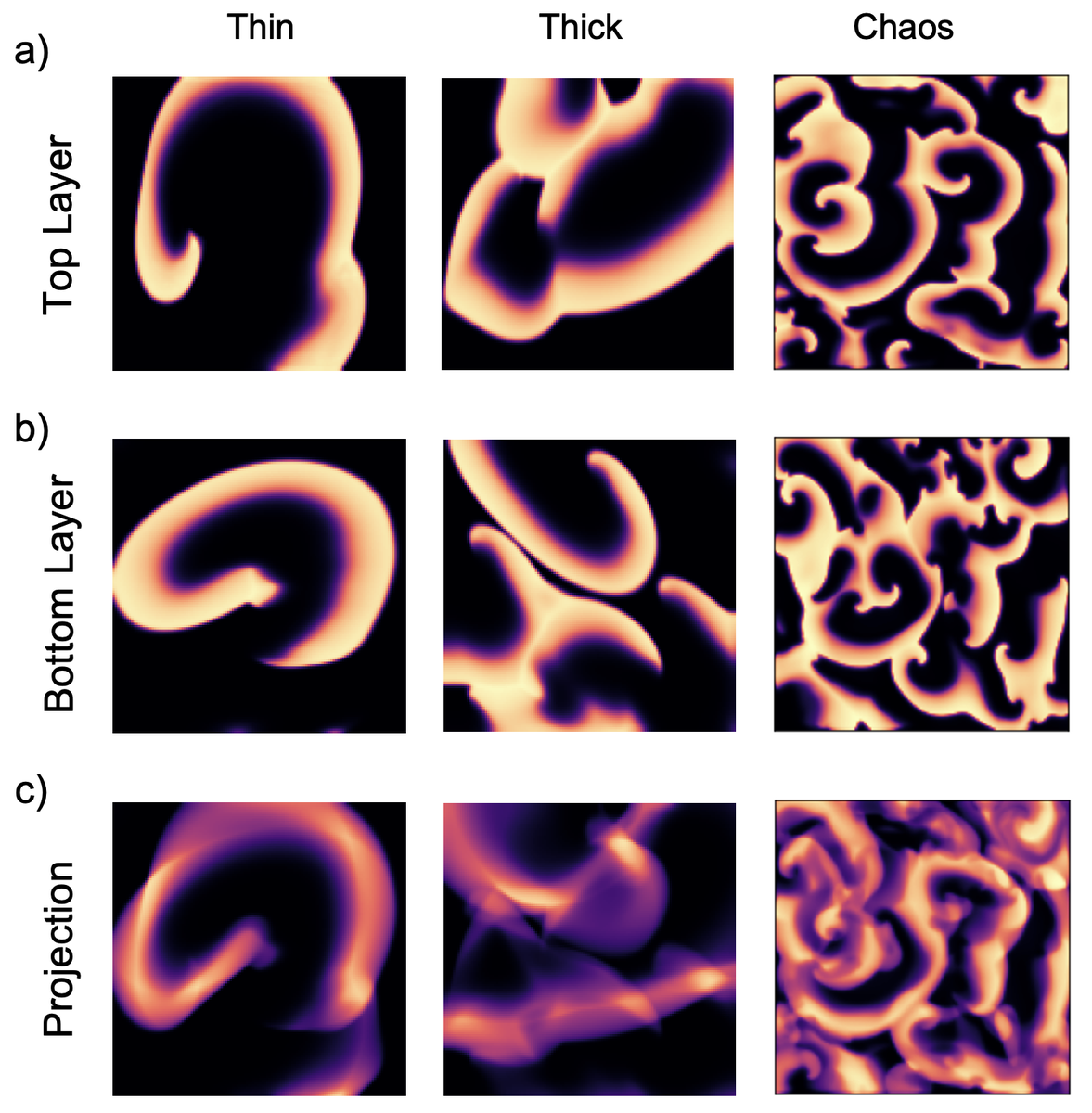}
  \caption{
    Different electrical wave patterns as seen in the a) top layer, b) bottom layer and c) projection of all layers of a three-dimensional bulk. 
    Bulk sizes are {$128 {\times} 128 {\times} 24$ voxels (left and right) and $128 {\times} 128 {\times} 40$ voxels (center), respectively}.
    {With increasing bulk thickness or smaller scroll waves}, the top and bottom layers are dissociated {because the dynamics become increasingly} three-dimensional.
    The projection is calculated for a given $(x,y)$-coordinate by averaging the $u$-values ($u \in [0, 1]$) along the depth ($z$-direction) of the bulk.
    {Data from left to right shown in Figs.~\ref{fig:ScrollWave}, \ref{fig:LargeScrollWave} and \ref{fig:scrollwavechaoslayers}, respectively.
    The left and center snapshots are from the `laminar' parameter regime, while the ones on the right are from the `turbulent' chaotic parameter regime.}
  }
  \label{fig:Figure03}
\end{figure}

\begin{figure}[htb]
  \centering
  \includegraphics[width=0.46\textwidth]{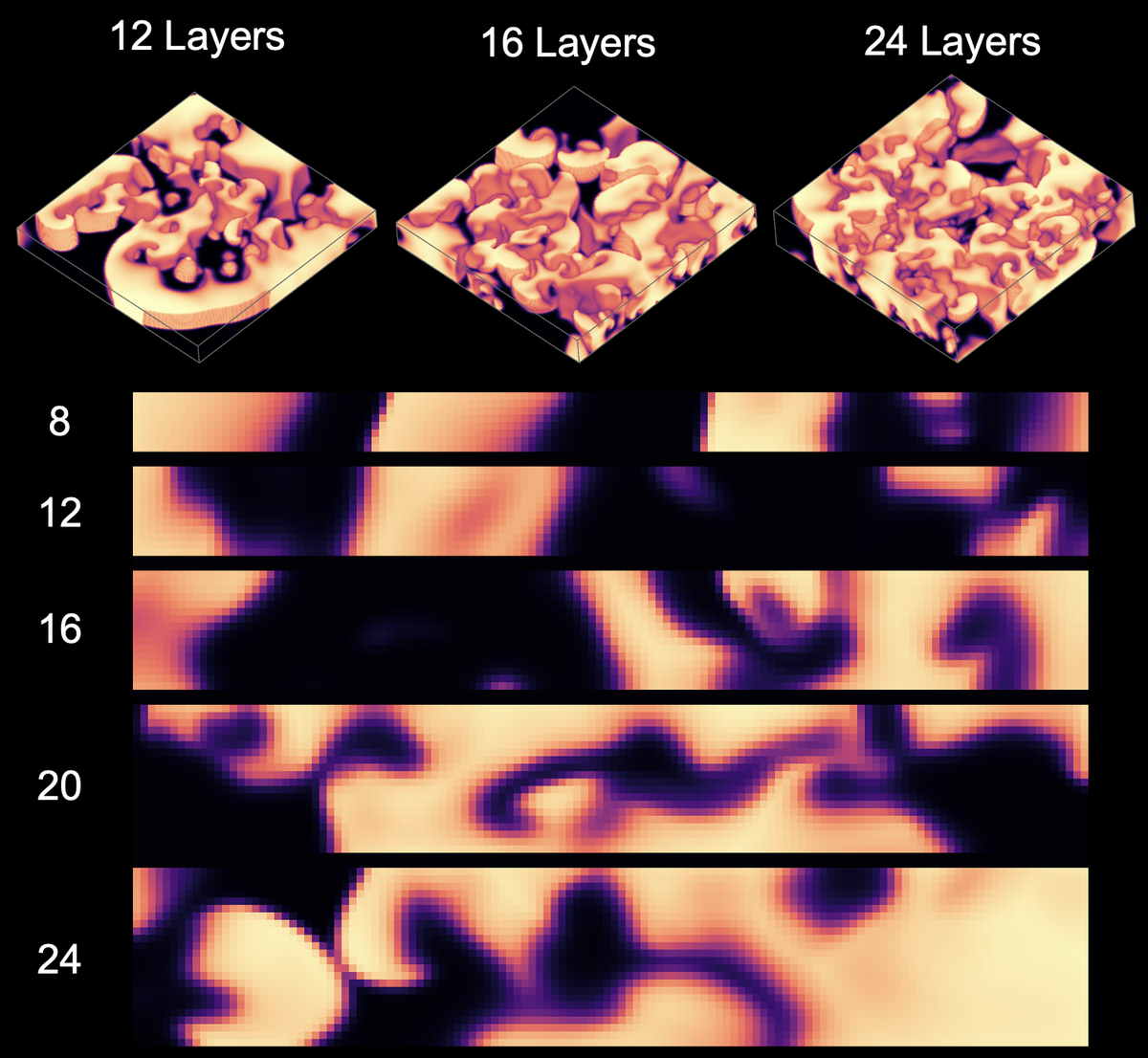}
  \caption{
Bulk thickness and transmurality of scroll wave dynamics. 
{'Turbulent'} scroll wave chaos with different bulk thicknesses $d_z = \{8,12,16,20,24\}$, also shown in corresponding cross-sections. 
The dynamics are quasi two-dimensional with $d_z=8$. 
Dissociation between top and bottom layers starts to emerge at $d_z=12$ as the dynamics become increasingly three-dimensional. 
At $d_z>12$ the dynamics are fully three-dimensional.
  }
  \label{fig:Figure04}
\end{figure}

\begin{figure*}[htb]
  \centering
  \includegraphics[width=0.98\textwidth]{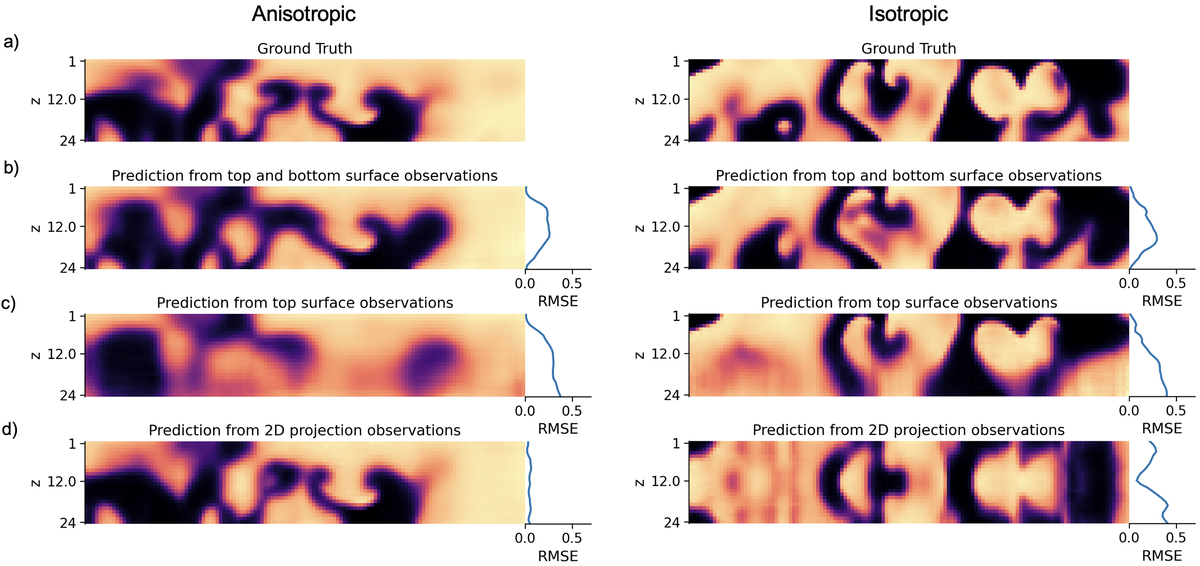}
  \caption{
    Predictions of {('turbulent')} electrical scroll waves in subsurface layers of anisotropic (left) and isotropic (right) bulk tissue with dimension $128 {\times} 128 {\times} 24$ voxels. 
    a) Ground-truth scroll wave dynamics (representative snapshots)
    b) Prediction in dual-surface mode analyzing the top and bottom layers of an opaque bulk tissue.
    c) Prediction in single-surface mode analyzing the top surface layer of an opaque bulk.
    d) Prediction analyzing the $z$-projection of the dynamics along its depth (or $z$-axis) in a transparent bulk.
    The depth-profile of the prediction error (RMSE: root mean squared error) along the $z$-axis is shown to the right of each prediction.
    The reconstruction is successful in anisotropic transparent media, but fails in isotropic transparent media, as shown in d).
    In opaque media, the reconstruction performs sufficiently well in dual-surface mode with larger errors emerging at midwall of the bulk, as shown in b).
    In transparent isotropic media, the subsurface prediction fails because the network is unable to infer the depth of the layers.
    Cross-sections intersect the bulk at its center. {All reconstruction with U-Net.}
  }
  \label{fig:ScrollWaveChaosCrosssections}
\end{figure*}

We implemented and tested deep neural networks, which each analyze a short temporal sequence of $5$ subsequent two-dimensional snapshots of electrical wave patterns $\tilde{u}_t(x, y)$ to reconstruct a single fully three-dimensional snapshot $u_t(x, y, z)$ of scroll wave {dynamics}:
\begin{eqnarray}
  \label{eq:bulkprediction}
  \left(\tilde{u}_1(x, y), \ldots, \tilde{u}_5(x, y)\right) \rightarrow u_1(x, y, z)\text{.}
\end{eqnarray}
{We found empirically that $5$ snapshots provide sufficient information about the dynamics, see also \cite{Lebert2021} or \cite{Christoph2020}.
We also tested reconstructing the dynamics with a series of $10$ snapshots and did not observe an improvement in performance compared to $5$ snapshots.
Consequently, we used $5$ snapshots by default.
Further, we found that the reconstruction accuracy does not depend on whether the network analyzes the current snapshots plus $4$ snapshots sampled in the past or in the future with respect to the current snapshot, or whether $2$ are sampled in the past and $2$ in the future, respectively.}
The two-dimensional snapshots, see Figs.~\ref{fig:Figure02}, \ref{fig:ScrollWaveMethod} and \ref{fig:Figure03}, are either
i) the top surface layer (single-surface mode, {Fig.~\ref{fig:Figure02}a}):
\begin{eqnarray}
  \tilde{u}_t(x, y) = u_t(x, y, 1)\text{,}
\end{eqnarray}
ii) both the top and bottom surface layer (dual-surface mode, {Fig.~\ref{fig:Figure02}b}):
\begin{eqnarray}
  \tilde{u}_t(x, y) = (u_t(x, y, 1), u_t(x, y, d_z))\text{,}
\end{eqnarray}
{iii)} both the electrical wave dynamics and the mechanical displacements $\vec{d} = (dx, dy)$ which occur in corresponding electromechanical simulations from the top surface layer (single-surface mode){, see Fig.~\ref{fig:Figure02}c)}:
\begin{eqnarray}
  \tilde{u}_t(x, y) = (u_t(x, y, 1), dx(x, y, 1), dy(x, y, 1))\text{,} \label{eq:displacements}
\end{eqnarray}
or {iv)} a projection of all $u$-values along the $z$-direction (depth) of the bulk {(Fig.~\ref{fig:Figure02}d)}:
\begin{eqnarray}
  \tilde{u}_t(x, y) = \frac{1}{d_z} \sum_{i=1}^{d_z} u_t(x, y, i)\text{.}
\end{eqnarray}
For the top+bottom case (iii) and the mechanical displacement case (iv), we interleaved the snapshots.
{
  The neural network architectures we chose require three-dimensional input samples ($(128, 128, 5)$ for case (i) and (ii)), whereas in cases (iii) and (iv) the sample shape is $(128, 128, 5, 3)$ and $(128, 128, 5, 2)$ respectively.
  We changed their shape to be three-dimensional by stacking the components: e.g. in case (iv) the resulting shape is $(128, 128, 10)$ using the top layers for the even indicies and bottom frames for the odd indices $(u_1(x, y, 1), u_1(x, y, d_z), u_2(x, y, 1) \ldots u_5(x, y, d_z))$.
}

We evaluated four different neural network architectures with basic and more intricate designs for the three-dimensional bulk prediction task (Eq.~\ref{eq:bulkprediction}).
While we primarily use a U-Net \cite{Ronneberger2015} architecture, we validate it against a simple {Encoder-Decoder} architecture, \mbox{TransUNet} \cite{TransUNet} and MIRNet \cite{mirnet}.
The {Encoder-Decoder} convolutional neural network (CNN) is similar to the architecture we previously used in \cite{Christoph2020,Lebert2021}.
It consists of an encoder stage where the spatial resolution is progressively decreased, a latent space, and a decoder stage where the spatial resolution is progressively increased back to the original resolution.
The encoding and decoding steps consist of three steps, in each two padded two-dimensional convolutional layers (2D-CNN) with filter size $3{\times}3$ and rectified linear unit \cite{Nair2010} (ReLU) activation are applied, followed by batch normalization \cite{Ioffe2015} and maxpooling (encoder) or upscaling (decoder), respectively.
{
The number of filters in 2D-CNN layers in order are 128, 128, 256, 256, 512, 512, 256, 256, 128 and 128.
}
The U-Net architecture is identical to the {Encoder-Decoder} CNN architecture, except that skip connections are added between the encoder and decoder stages (see \cite{Ronneberger2015}).
The TransUNet combines the U-Net architecture with self-attention mechanisms of Transformers \cite{AttentionIsAllYouNeed} in its the latent space.
The MIRNet architecture is different from the other evaluated architectures, as it contains parallel multi-resolution branches with information exchange, as well as spatial and channel attention mechanisms \cite{mirnet}.
It aims at maintaining spatially-precise high-resolution representations through the entire network, while simultaneously receiving strong contextual information from the low-resolution representations.
For all neural network architectures we use the generalized Charbonnier loss function \cite{Bruhn2005,Barron2019}:
\begin{eqnarray}
  l(u, \hat{u}) = \sqrt{(\hat{u} - u)^2 + \epsilon^2}\text{,}
\end{eqnarray}
where $u$ is the three-dimensional ground truth, $\hat{u}$ the prediction and we choose $\epsilon = 0.001$.
The Charbonnier loss function behaves like L2 loss (mean squared error) when $u \approx \hat{u}$ and like L1 loss (mean absolute error) otherwise.
We evaluated the accuracy of the predictions on the evaluation datasets with the root mean squared error (RMSE) on each $z$-axis layer:
\begin{eqnarray}
  \text{RMSE}(z) = \sqrt{\frac{1}{N}\sum_{x,y,t}(\hat{u}(x,y,z,t) - u(x,y,z,t))^2}\text{.}
\end{eqnarray}

To validate our findings, we studied if a neural network can accomplish a simpler task than the three-dimensional prediction: estimate the depth $d_z$ of the simulation bulk from 5 two-dimensional observations. We tested both a depth regression and a depth classification neural network, which each predict the depth $d_z$ of the simulation bulk 
\begin{eqnarray}
  \left(\tilde{u}_1(x, y), \ldots, \tilde{u}_5(x, y)\right) \rightarrow d_z\text{.}
\end{eqnarray}
The depth regression network predicts the depth $d_z$ as a continuous value, while the depth classification network predicts the depth $d_z$ as one of $\{8, 12, 16, 20, 24, 28, 32\}$.
For this task we used the encoder part from the {Encoder-Decoder} architecture, followed by a global average pooling layer, two dense layers with $1024$ filters with batch normalization and ReLU activation, and ultimately an output dense layer with one filter (for regression) or seven filters (for classification).
For the depth classification neural network we used a categorical cross-entropy loss function with a softmax activation function for the last layer, and for the depth regression network mean squared error as loss function and ReLU as activation function.
The datasets for the depth estimation was generated from the bulk prediction task datasets. 
We used $4{,}000$ random samples for each depth for the training dataset and $500$ samples for the evaluation dataset (in total $28{,}000$ training samples and $3{,}000$ evaluation samples). 
\par
The networks were trained using the Adam \cite{Adam} optimizer with a learning rate of $10^{-3}$ for the bulk prediction tasks and $10^{-5}$ for the depth regression and classification task for $20$ epochs. 
We used a batch size of $32$ for the Encoder-Decoder and U-Net architectures and a batch size of $4$ for TransUNet and MIRNet.
All neural network models were implemented in Tensorflow \cite{tensorflow2015-whitepaper} using Keras \cite{chollet2015keras}.
Training and reconstructions were performed on NVIDIA RTX A5000 graphics processing units (GPUs).

\begin{table}[htb]
  \begin{tabular}{@{}lcc@{}}\toprule
    Model                & Parameters        & Training Time      \\\midrule
    Encoder-Decoder      & $6{,}829{,}309$   & $44\,\text{min}$   \\
    U-Net                & $8{,}278{,}168$   & $55\,\text{min}$   \\
    TransUNet            & $406{,}899{,}608$ & $19\,\text{hours}$ \\
    MIRNet               & $145{,}358{,}026$ & $17\,\text{hours}$ \\\midrule
    $d_z$ Regression     & $426{,}593$       & $5\,\text{min}$    \\
    $d_z$ Classification & $432{,}743$       & $5\,\text{min}$    \\\bottomrule
  \end{tabular}
  \caption{Different neural network architectures used in this study and their respective number of trainable parameters and training times for 20 epochs. Training was performed on a single NVIDIA RTX A5000 GPU.}
  \label{tab:network_types}
\end{table}

\section{Results}
\begin{figure*}[htb]
   \centering
   \includegraphics[width=0.95\textwidth]{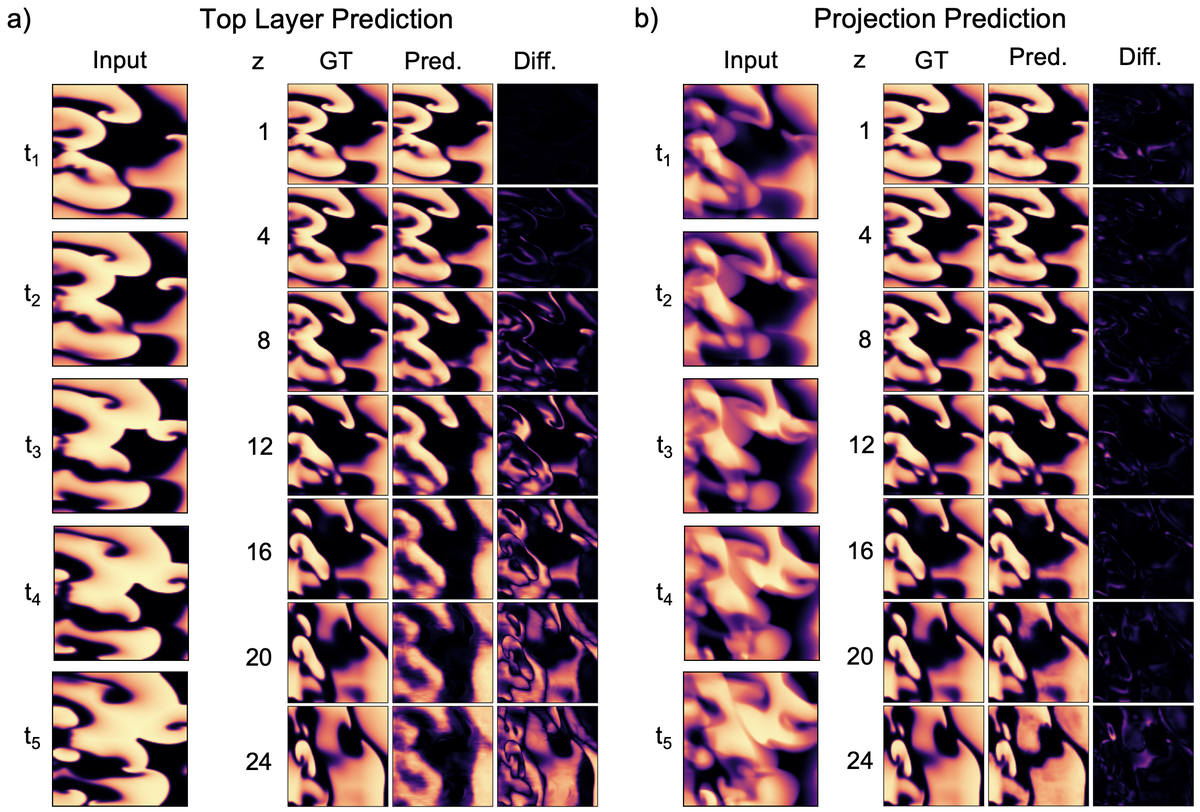}
   \caption{
 Predictions of electrical scroll waves within subsurface layers of a bulk-shaped `turbulent' anisotropic excitable medium from observing either a) the top layer of the bulk or b) the projection of the three-dimensional wave pattern in a transparent bulk along its depth (or z-axis). The bulk's dimensions are $128 {\times} 128 {\times} 24$ voxels {and predictions were performed with the U-Net architecture}.
 Predictions are shown for the 24 layers along the $z$-axis of the bulk, where the first layer is the top layer and the $24^{th}$ layer is the bottom layer.
 First row: The five two-dimensional frames ($t_1,\ldots,t_5$) which are the input for the neural network prediction.
 Second row: Ground truth (GT) electrical excitation wave pattern within cross-sectional layers (1-24), of which layers 2-24 cannot be observed. Note that the pattern changes from layer to layer throughout the bulk.
 Third row: Prediction of the current cross-sectional layer (1-24) by the neural network.
 Fourth row: Absolute difference per voxel between prediction and ground-truth.
   }
   \label{fig:scrollwavechaoslayers}
\end{figure*}

\begin{figure*}[htb]
  \centering
  \includegraphics[width=0.98\textwidth]{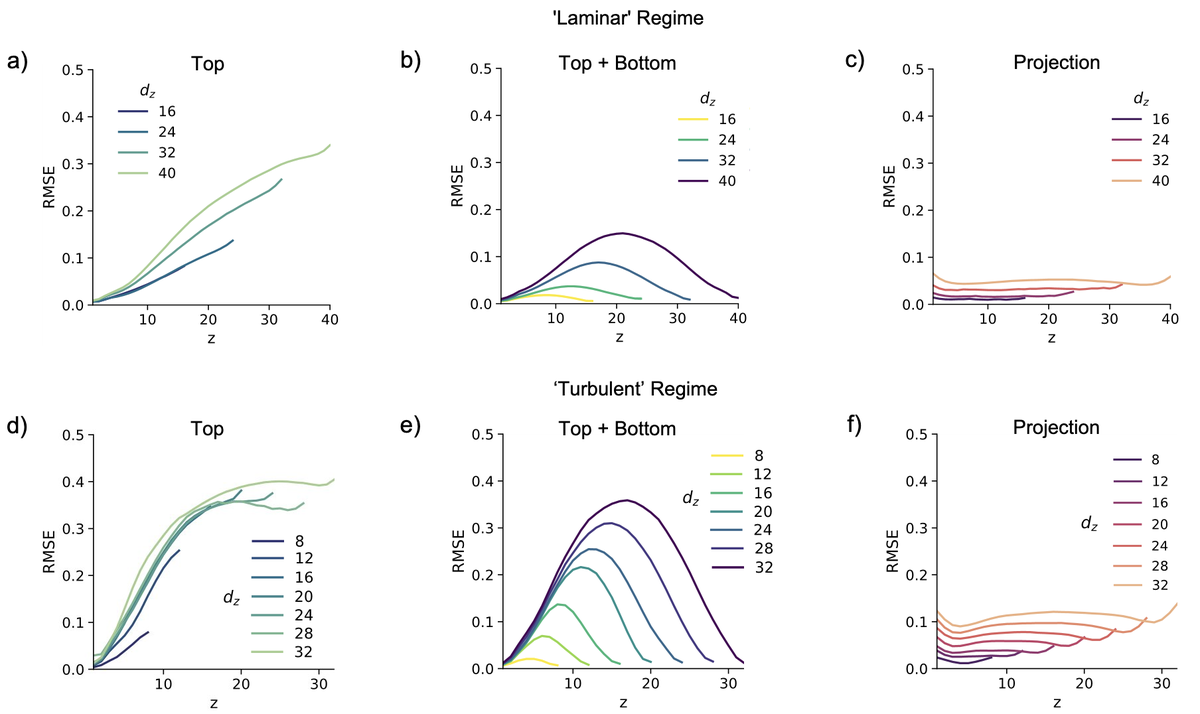}
  \caption{
    Average reconstruction error {over bulk depth} (RMSE: root mean squared error along $z$-axis) in a,b,d,e) opaque or c,f) transparent excitable media with anisotropy (with varying bulk depths of $d_z \in \{8, 12, \ldots , 40\}$). {All reconstructions were performed with U-Net.
    a-c) `Laminar' scroll wave dynamics as shown in Figs.~\ref{fig:ScrollWave} and \ref{fig:LargeScrollWave}: 
    a) Single-surface mode, see also Figs.~\ref{fig:ScrollWave}b) and \ref{fig:LargeScrollWave}b).
    b) Dual-surface mode.
    c) Projection, see also Figs.~\ref{fig:ScrollWave}c) and \ref{fig:LargeScrollWave}c).
    `Turbulent' scroll wave chaos as shown in Fig.~\ref{fig:Figure04}:
    d) Single-surface mode, see also Figs.~\ref{fig:ScrollWaveChaosCrosssections}c) and \ref{fig:scrollwavechaoslayers}a).
    e) Dual-surface mode, see also Figs.~\ref{fig:Figure02}b) and \ref{fig:ScrollWaveChaosCrosssections}b).
    f) Projection, see also Figs.~\ref{fig:ScrollWaveChaosCrosssections}d) and \ref{fig:scrollwavechaoslayers}b).}
    An error of $0.1$ (RMSE) corresponds to a mean absolute error (MAE) of about $5\%$.
    In opaque excitable media, the reconstruction error increases approximately linearly with depth.
    In transparent media with anisotropy the error remains flat and below $0.1$.
    We trained separate U-Net neural networks for each combination.
    {See also Fig.~S1 for a comparison with isotropic excitable media and Fig.~S2 for a comparison of the deep learning-based reconstruction with a naive reconstruction in which the top layer is simply repeated in each following layer.}
  }
  \label{fig:plots-scrollwavechaos}
\end{figure*}

Using deep convolutional neural networks, it is possible to reconstruct three-dimensional scroll wave {dynamics inside an excitable medium when the medium's thickness is not much thicker than the scroll wave, see Fig.~\ref{fig:ScrollWave} and section \ref{sec:results:scrollwavesize}.
Reconstructions become increasingly difficult in thicker excitable media or with smaller scroll waves and more complicated dynamics, see Figs.~\ref{fig:LargeScrollWave}, \ref{fig:scrollwavechaoslayers}a) and sections \ref{sec:results:scrollwavesize}-\ref{sec:results:transparency}.
However, complicated scroll wave chaos can be reconstructed in transparent anisotropic excitable media, see Figs.~\ref{fig:ScrollWaveChaosCrosssections}d), \ref{fig:scrollwavechaoslayers}b), \ref{fig:plots-scrollwavechaos}c,f), or with dual-surface observations in thinner opaque excitable media, see Figs.~\ref{fig:LargeScrollWave}c) and \ref{fig:ScrollWaveChaosCrosssections}b) and section \ref{sec:results:singlevsdualsurface}.}

{
\subsection{Medium Thickness vs. Scroll Wave Size} \label{sec:results:scrollwavesize}
The thickness of the excitable medium and the size of the scroll waves with respect to the thickness of the medium determine to what extend and with which accuracy scroll wave dynamics can be reconstructed.
To test the reconstruction performance with different size ratios, we simulated `laminar' scroll waves, see Figs.~\ref{fig:ScrollWave} and \ref{fig:LargeScrollWave}, and `turbulent' scroll wave chaos, see Figs.~\ref{fig:Figure04}, \ref{fig:ScrollWaveChaosCrosssections}c) and \ref{fig:scrollwavechaoslayers}a), in bulks with varying thicknesses.
While the reconstructions from single surface observations are very accurate for the `laminar' scroll wave dynamics in the thinner bulk (with thickness $d_z=24$) shown in Figs.~\ref{fig:ScrollWave}a,b) and in Supplementary Video 1, reconstructions become increasingly inaccurate with increasing thickness and prediction depth. 
The reconstructions of the `laminar' scroll wave dynamics in the thicker bulk (with thickness $d_z=40$), shown in Fig.~\ref{fig:LargeScrollWave}a,b) and in Supplementary Video 2, exhibit artifacts towards the bottom half of the bulk. 
Accordingly, the vortex filaments computed from the predicted scroll wave dynamics (red) exhibit substantial mismatches compared to the gound-truth vortex filaments (gray) in the lower half of the bulk, see Fig.~\ref{fig:LargeScrollWave}d).
With increasing bulk thickness, the three-dimensional character and complexity of the wave dynamics increases, which is reflected by the dissociation of the top and bottom layers, see Figs.~\ref{fig:Figure03} and \ref{fig:Figure04}, and also by the various orientations of the vortex filaments in Fig.~\ref{fig:LargeScrollWave}d).
The degree of dissociation and the average scroll wave size relative to the medium's thickness ultimately determine the prediction accuracy at deeper layers.
The 'horizon' up to which the predictions are successful appears to be approximately one scroll wavelength, also compare the bottom half of the thick bulk with `laminar' scroll wave dynamics in Fig.~\ref{fig:LargeScrollWave}b) to the bottom half of the thinner bulk with `turbulent' scroll wave chaos in Figs.~\ref{fig:ScrollWaveChaosCrosssections}b) and \ref{fig:scrollwavechaoslayers}a).
In general, the reconstructions become increasingly difficult the deeper one aims to predict in opaque excitable media, and they do not succeed beyond the first layer of scroll waves.}

\begin{figure}[htb]
  \centering
  \includegraphics[width=0.48\textwidth]{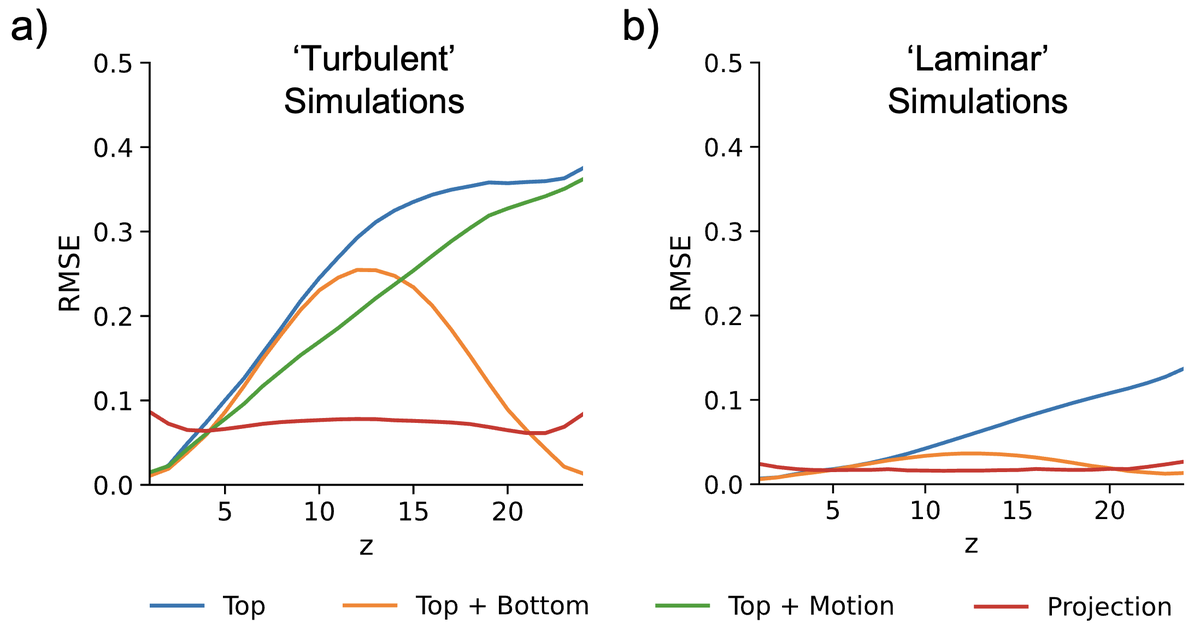}
  \caption{
    {
    Comparison of reconstruction errors obtained with different imaging configurations in opaque anisotropic excitable medium (with thickness $d_z=24$ layers) with a) `turbulent' scroll wave chaos and b) `laminar' scroll wave dynamics. 
    In the different configurations, the network (U-Net) analyzes i) in single-surface mode the electrics on the top layer (blue), ii) in single-surface mode the electrics and motion on the top layer (green), iii) in dual-surface mode the electrics on both top and bottom layers (orange), and iv) the $z$-projection of the three-dimensional electrics (red).
    All reconstruction errors were calculated per depth as root mean squared error (RMSE).}}
  \label{fig:plotsNewParams}
\end{figure}

\begin{figure}[htb]
  \centering
  \includegraphics[width=0.48\textwidth]{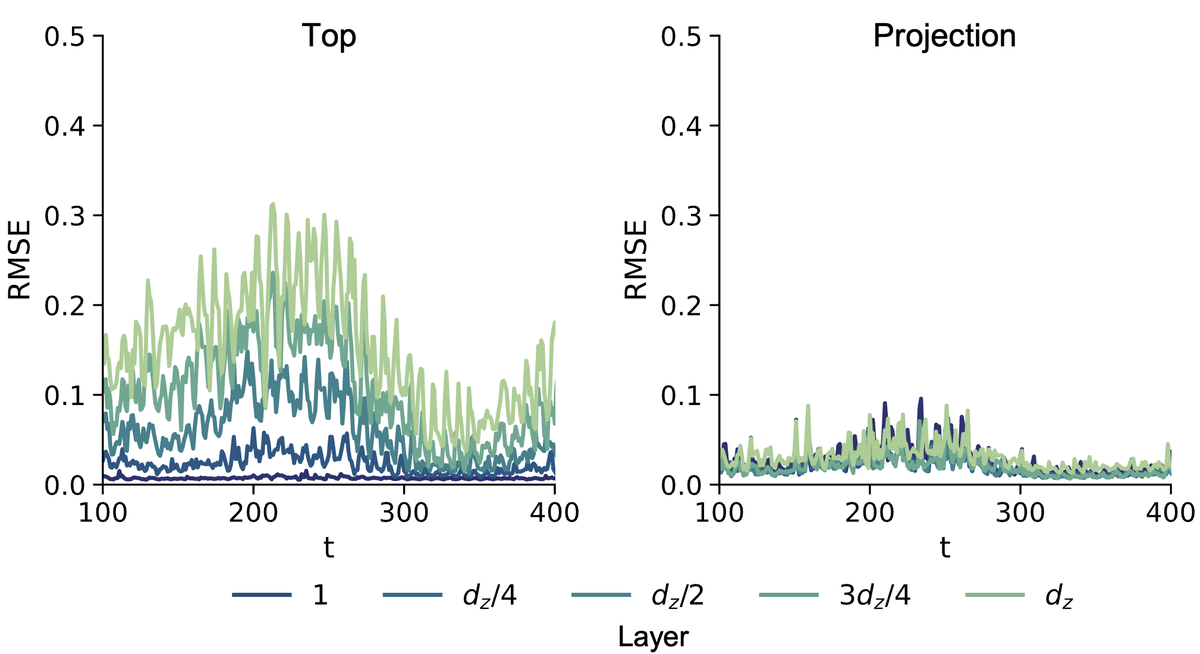}
  \caption{
{Average reconstruction error per layer (depth) over time in an anisotropic excitable medium with thickness $d_z=24$ shown for $5$ different depths.
  Left: In opaque media and single-surface mode, the prediction error increases and fluctuates more with increasing depth.
  Right: In transparent media, the prediction error stays small throughout the different depths.
  Both plots derived for the `laminar' scroll wave dynamics as shown in Fig.~\ref{fig:ScrollWave}.}}
  \label{fig:temporal_evolution}
\end{figure}

{
Correspondingly, the plots in Fig.~\ref{fig:plots-scrollwavechaos}a,d) show how the prediction error increases in opaque excitable media with increasing depth $z$ (in bulks with different depths $d_z=8,12,\ldots,40$) with `laminar' and `turbulent' scroll wave dynamics, respectively. 
The error increases approximately linearly with increasing depth and increases faster with thicker bulks and faster with `turbulent' than with `laminar' scroll wave dynamics.
We compared the error profiles obtained with the deep learning-based reconstruction with a naive reconstruction in which the top layer is simply repeated in each following layer in Fig.~S2 in the Supplementary Information.
The naive reconstruction produces significantly steeper error curves with both `laminar and `turbulent' scroll wave dynamics. 
The curves in Fig.~\ref{fig:plots-scrollwavechaos}a,d) indicate that the average reconstruction accuracies in Fig.~\ref{fig:ScrollWave}b) and Fig.~\ref{fig:LargeScrollWave}b) are approximately $0.05$ and $0.15$ (RMSE) at midwall, respectively, while the deepest layers shown in Fig.~\ref{fig:LargeScrollWave}b) yield reconstruction errors in the order of $0.2-0.5$ (RMSE).
By comparison, the worst average reconstruction error at the maximal depth of the `laminar' scroll wave shown in Fig.~\ref{fig:ScrollWave}b) is about $0.1$ (RMSE), see blue curve in Fig.~\ref{fig:plotsNewParams}b).
Note that a root mean squared error of $0.1$ (RMSE) corresponds to about $0.05$ mean absolute error (MAE).
The average single-surface reconstruction error for the `laminar' scroll wave in Fig.~\ref{fig:ScrollWave}b) is better than $95\%$.
Overall, the reconstruction error fluctuates moderately over time, remains small at smaller depths, and increases as the reconstruction error increases with larger depths, see Fig.~\ref{fig:temporal_evolution}.
Supplementary Videos 1-4 demonstrate the reconstructions for `laminar' and 'turbulent scroll' wave dynamics and give an impression of the temporal stability of the reconstructions.}

{
\subsection{Single-Surface vs. Dual-Surface Observations} \label{sec:results:singlevsdualsurface}
Low prediction depths in opaque excitable media can be overcome by analyzing both the top and bottom surface layers in dual-surface mode rather than single-surface mode, respectively.
Figs.~\ref{fig:LargeScrollWave}c) and \ref{fig:ScrollWaveChaosCrosssections}b) demonstrate how the reconstruction improves in a thick bulk with the `laminar' scroll wave and in a thin bulk with `turbulent' scroll wave chaos,  respectively.
The vortex filaments (red) in the thick bulk in Fig.~\ref{fig:LargeScrollWave}c) match the gound-truth vortex filaments (gray) much better in dual- than in single-surface mode.
The plots in Fig.~\ref{fig:plots-scrollwavechaos}b,e) and \ref{fig:plotsNewParams} show how the profile of the reconstruction error changes in dual-surface mode.
Surprisingly, the network does not appear to benefit from the additional information from both surfaces of the bulk in dual-surface mode with scroll wave chaos: the steep linear increase in the error perists on both sides and the network is not able to significantly reduce the error at midwall, see Fig.~\ref{fig:plots-scrollwavechaos}b).
Nevertheless, it is possible to slightly reduce the error at midwall when reconstructing larger scroll waves, see Figs.~\ref{fig:LargeScrollWave}c), \ref{fig:plots-scrollwavechaos}b) and \ref{fig:plotsNewParams}b).
The data suggests that it could be possible to reconstruct scroll waves in the heart using epi- and endocardial optical mapping recordings, if the scroll wavelength is not much shorter than the thickness of the ventricular wall.}

\subsection{Transparent vs. Opaque Excitable Media}\label{sec:results:transparency}
{While it is challenging to reconstruct scroll wave dynamics or scroll wave chaos in thicker opaque excitable media, the reconstructions succeed in transparanet excitable media of any thickness (that we tested). 
However, the reconstructions only succeed under the condition that the excitable media are anisotropic.}
Figs.~\ref{fig:ScrollWaveChaosCrosssections} and \ref{fig:scrollwavechaoslayers} show a comparison of reconstructions obtained in an opaque and transparent excitable medium, respectively.
Fig.~\ref{fig:ScrollWaveChaosCrosssections} shows cross-sections along the depth of the bulk ($z$-direction) whereas Fig.~\ref{fig:scrollwavechaoslayers} shows layers parallel to the surface in the $x$-$y$ plane of the bulk.
Comparing Fig.~\ref{fig:ScrollWaveChaosCrosssections}c,d) and Fig.~\ref{fig:scrollwavechaoslayers}a,b), it becomes immediately apparent that it is possible to obtain highly accurate reconstructions of scroll wave chaos in transparent, anisotropic excitable media, while it would not be possible to obtain similar reconstructions in opaque media (see also additional plots in Fig.~S1 with isotropy).
Fig.~\ref{fig:plots-scrollwavechaos}c) shows that the reconstruction error stays small transmurally throughout the entire bulk, if the excitable medium is transparent and anisotropic.
The prediction error is $<0.1$ (RMSE) with {various} bulk depths ($d_z=8-32$) and with the `laminar' and `turbulent' scroll wave dynamics.
Importantly, as can be seen in the right panel in Fig.~\ref{fig:ScrollWaveChaosCrosssections}d), the reconstruction completely fails in isotropic transparent excitable media, see also Fig.~S1 in the Supplementary Information.
We discuss the effect of anisotropy onto the prediction in more detail in the next section and in the discussion.

\subsection{Anisotropy}
\label{sec:results:aniso}
While ventricular muscle tissue is highly anisotropic {(orthotropic muscle fiber organization), the Belousov-Zhabotinsky chemical reaction is isotropic.
Both systems exhibit scroll waves, but the scroll wave morphology can be very different in anisotropic versus isotropic excitable media.
In anisotropic media, scroll waves are elongated in fiber direction as they propagate faster along the fiber direction. 
This phenomenon can often be observed in optical mapping recordings.}
In the simulated anisotropic bulk, the waves are elongated differently at different depths, which is presumably {why the reconstructions succeed in transparent excitable media as shown in Fig.~\ref{fig:plots-scrollwavechaos}c,f).}
By contrast, in isotropic excitable media the scroll waves are similarly shaped throughout the bulk, and therefore the network cannot distinguish scroll waves closer to the surface from scroll waves deeper in the bulk, see also discussion and Fig.~S1 in the Supplementary Information.
Anisotropy does not affect the reconstructions in opaque excitable media, as the reconstruction does not rely on depth information.

{
\subsection{Analyzing Surface Deformation}
If the network analyzes the mechanical deformation of the surface in addition to the excitable wave patterns visible on the same surface, the reconstruction improves slightly, see plot 'Top + Motion' in Fig.~\ref{fig:plotsNewParams}a).
We tested this behavior with `turbulent' scroll wave chaos, the U-Net architecture and the single-surface configuration shown in Fig.~\ref{fig:Figure02}c), and found that the reconstruction improves slightly, but not substantially.
This is a surprising finding, because deformation on the surface may also occur due to contractile activity within the bulk.
The reconstruction error does not increase as steeply with increasing depth as when analyzing electrics in single-surface mode alone.
The reconstruction accuracy improves by roughly 25\% at midwall (with a bulk thickness of $d_z=24$ layers).
We found that there was no significant difference between analyzing only two-dimensional in-plane displacements $\vec{u} = (u_x, u_y)$ with $x$- and $y$-components versus three-dimensional displacements with also a $z$-component, see also eq. \eqref{eq:displacements} in section \ref{sec:methods:deeplearning}.
We cannot exclude that the latter finding is specific to our methodology and the mechanical boundary conditions that we used in the simulations.
}

\begin{figure}[htb]
  \centering
  \includegraphics[width=0.48\textwidth]{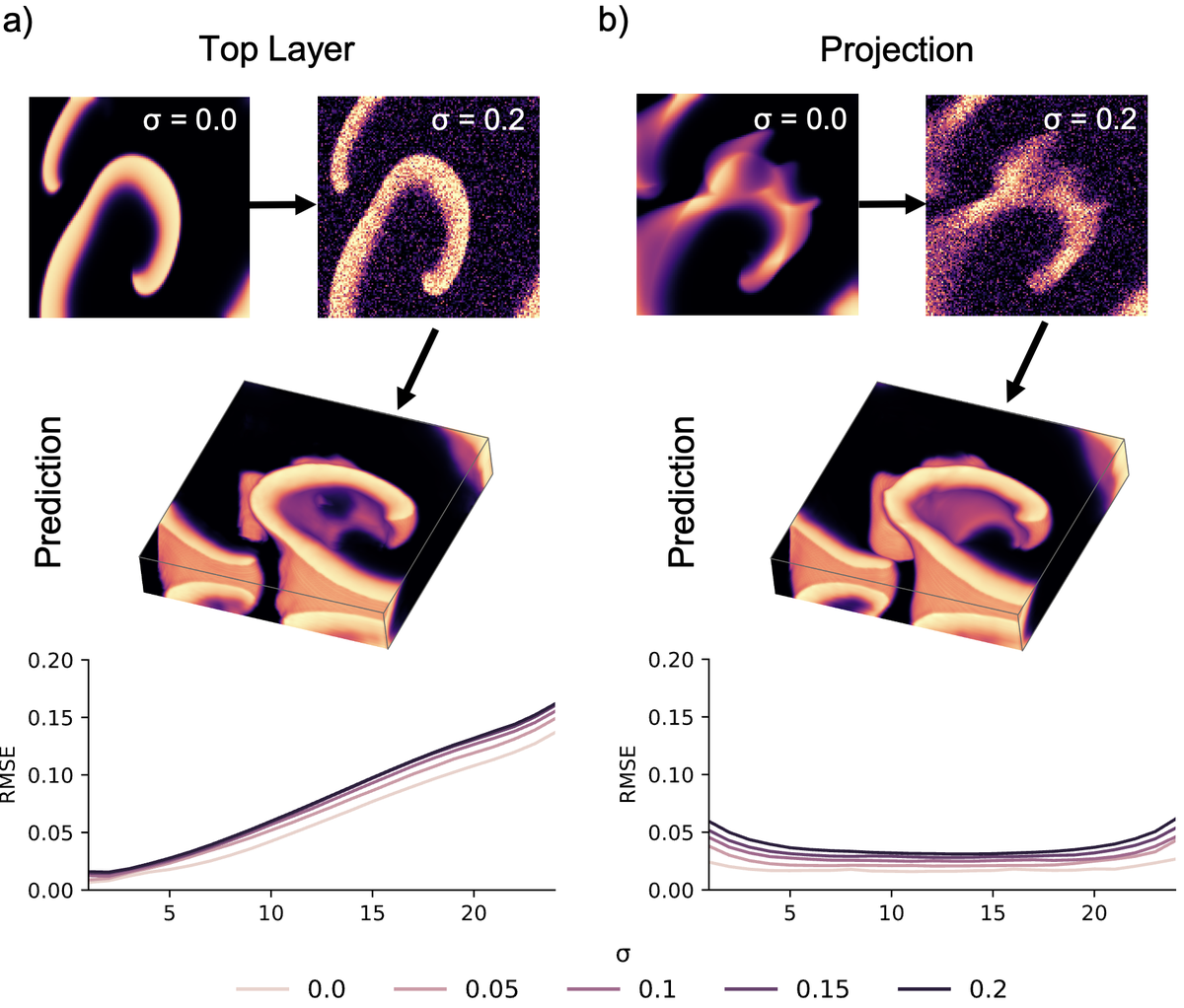}
  \caption{{Noise does not pose a limitation for the deep learning-based reconstructions (performed with U-Net on `laminar' scroll wave dynamics).
  Reconstructions succeed in the presence of various noise levels ($\sigma = 0.05, \ldots, 0.2$) in both a) opaque and b) transparent anisotropic excitable media.
  Top: Gaussian noise with standard deviation $\sigma$ was added onto the input images.
  Bottom: Error profiles along depth increase only slightly (light pink curve: $\sigma=0.0$, black curve: $\sigma=0.2$).
  }}
  \label{fig:Noise}
\end{figure}

\subsection{{Noise}}

{
Reconstructions can be performed with noise, see Fig.~\ref{fig:Noise} and Supplementary Video 5, if the network was previously trained with noise, similarly as described in \cite{Lebert2021} and \cite{Christoph2020}.
The noise can be present in either the surface observations or projections. 
Fig.~\ref{fig:Noise}a,b) shows reconstructions of scroll wave dynamics with noise in an opaque and a transparanet anisotropic excitable medium (thickness: $d_z=24$ layers), respectively. 
In the opaque bulk in single-surface mode, the slope of the reconstruction error is slightly steeper with noise than without. 
In the transparent bulk, the error profile remains flat and stays below $0.1$ (RMSE) with noise. 
We tested this bevaior with Gaussian noise and noise levels of up to $\sigma=0.2$, shown in Fig.~\ref{fig:Noise} (top right in each panel).
}

\begin{figure}[htb]
  \centering
  \includegraphics[width=0.48\textwidth]{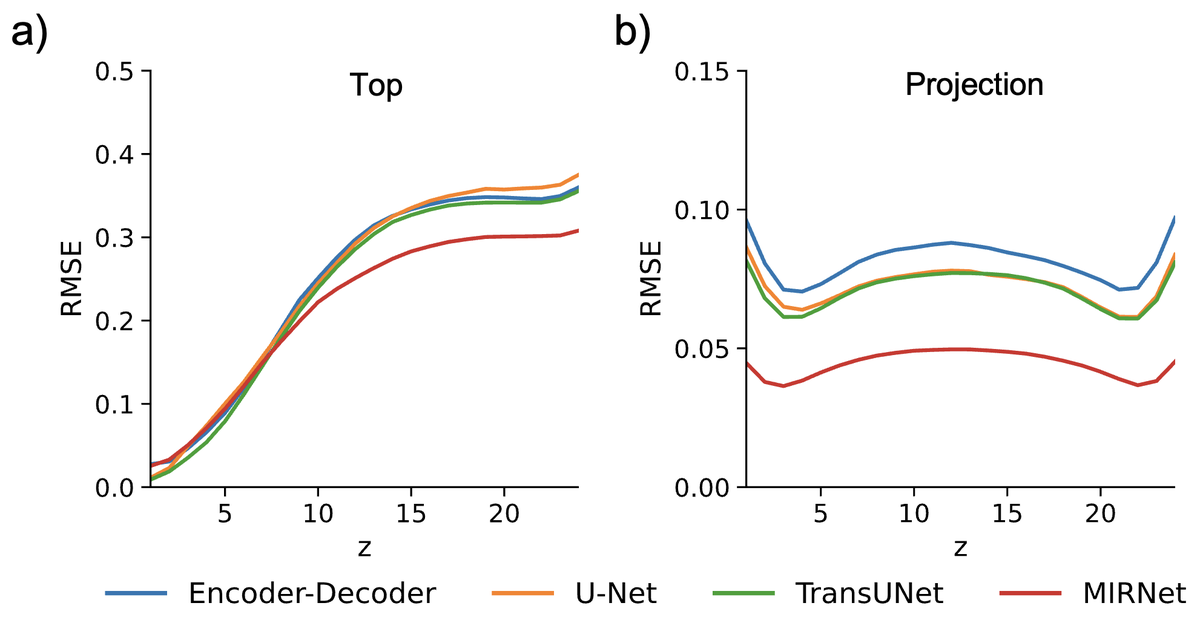}
  \caption{
    {Reconstruction errors obtained with different neural network architectures for `turbulent' scroll wave chaos in a bulk with thickness of $d_z = 24$.
    a) Steep increase of reconstruction error with all networks (Encoder-Decoder, U-Net, TransUNet, MIRNet) in opaque excitable media from a single surface (top).
    b) Low and relatively flat reconstruction errors (below $10\%$) with all networks in transparent anisotropic excitable media (projection).
    MIRNet performs slightly better than the other networks.
    All reconstruction errors stated as root mean squared error (RMSE).}
    }
  \label{fig:plots2}
\end{figure}

\subsection{Network Types}

We tested several deep neural network architectures (basic Encoder-Decoder, U-Net, TransUNet and MIRNet, see also section \ref{sec:methods:deeplearning}) on the {`turbulent'} scroll wave chaos prediction task and found that the prediction behavior is very similar across the different architectures, see {Fig.~\ref{fig:plots2}}.
We observed that the MIRNet architecture produces the lowest reconstruction error, while Encoder-Decoder, U-Net, and TransUNet all have similar but slightly higher reconstruction errors.
In opaque excitable media, the prediction error (RMSE: mean root squared error) rises linearly and steeply with deeper layers equally with all networks, as seen for the single-surface reconstructions shown in Fig.~\ref{fig:plots2}a) for anisotropic scroll wave chaos in a bulk with thickness $d_z=24$.
The prediction error saturates equally with all networks at depths $d_z>10$ where they produce maximal prediction errors of about $0.3$ (RMSE).
MIRNet provides a slightly lower maximal error than the other networks.
All networks achieve small prediction errors of $<0.1$ (RMSE) in transparent excitable media with anisotropy, as seen for the projection reconstructions shown in Fig.~\ref{fig:plots2}b) for anisotropic scroll wave chaos in a bulk with thickness $d_z=24$.
MIRNet provides the lowest error of less than $0.05$ (RSME), whereas the other networks produce errors ranging between $0.06-0.1$ (RMSE).
All networks produce the same characteristic error profile.
Note that a RMSE of $0.1$ corresponds to a mean absolute error (MAE) of about $5\%$. 
Therefore, all networks achieve reconstruction accuracies of greater than $95\%$ in transparent excitable media.
MIRNet even achieves reconstruction accuracies in the order of $97\%-98\%$.

As described in section \ref{sec:methods:deeplearning}, U-Net differs from the Encoder-Decoder architecture in the inclusion of long skip connections, while TransUNet is a U-Net with a Transformer as the latent space.
MIRNet is different in that it has multi-resolution branches with information exchange as well as self-attention mechanisms.
The training times for 20 epochs and the number of trainable parameters for each network architecture are listed in Table~\ref{tab:network_types}.
Given that TransUNet and MIRNet provided either no or incremental improvements in prediction accuracy, while requiring significantly longer time to train, we primarily used U-Net in this study. 
U-Net required about an hour to train, while being competitive with the accuracy of MIRNet, which required almost a full day to train.
All other results in Figs.~\ref{fig:ScrollWave}-\ref{fig:scrollwavechaoslayers} were obtained with the U-Net architecture, if not stated otherwise.
We tested several U-Net sizes: a small model with 0.5M parameters, a medium network with 2M parameters and a large model with 8M parameters, determined that larger models perform significantly better, and subsequently used the largest model.
In some circumstances U-Net and TransUNet exhibited a significantly better reconstruction performance than the Encoder-Decoder network, but we did not observe significant differences in accuracy between U-Net and TransUNet.

\subsection{Depth Estimation}

\begin{figure}[htb]
  \centering
  \includegraphics[width=0.48\textwidth]{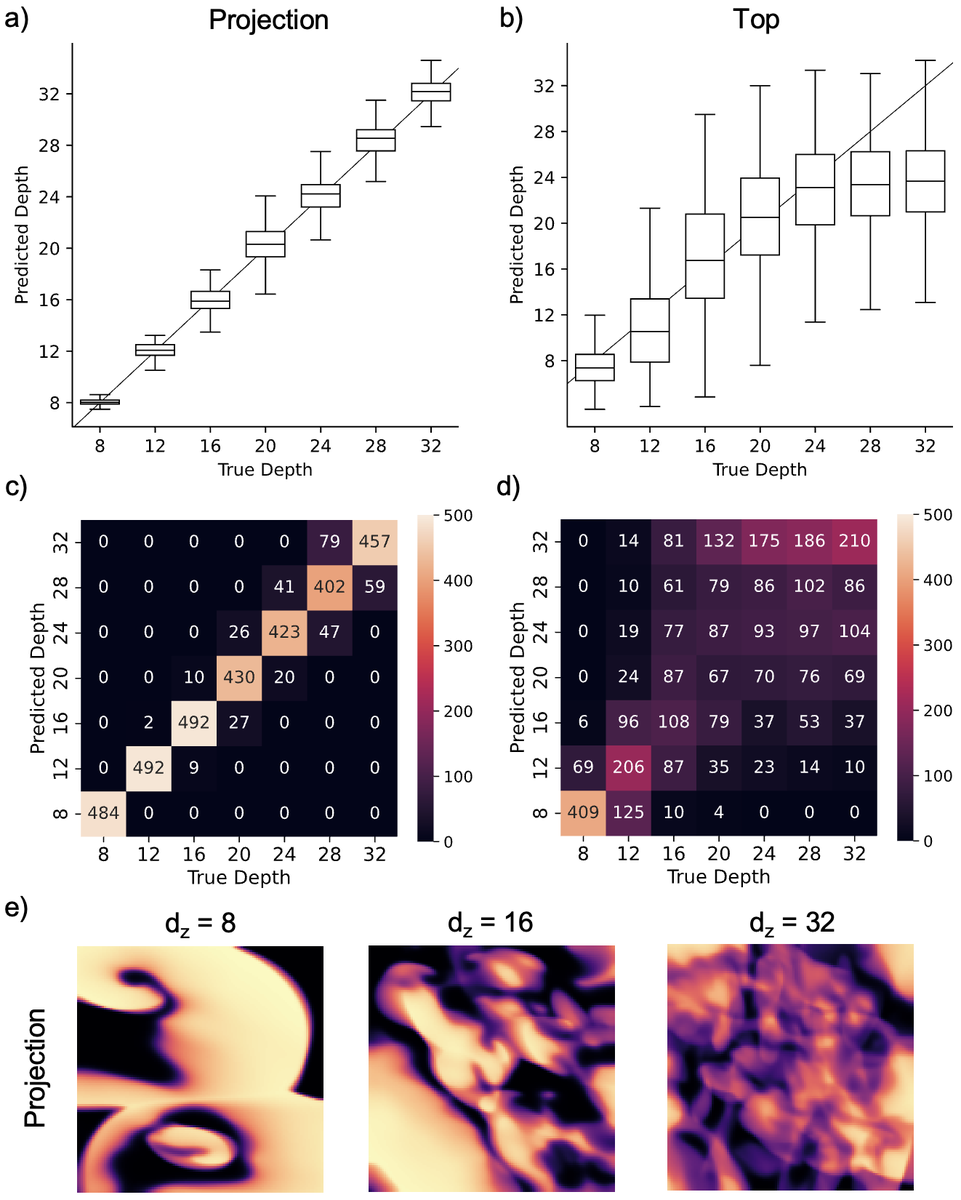}
  \caption{
    Prediction of bulk thickness from surface observations of scroll wave chaos in a,c) transparent bulk medium with projection observations and b,d) opaque bulk medium with top surface observations.
    {The bulk thickness was predicted using either a,b) a regression (black line shows ideal prediction $d_{\text{prediction}}=d_{\text{true}}$) or c,d) a classification neural network, respectively (all in anisotropic media).}
    In transparent media, the thickness can be predicted from observations as shown in Fig. \ref{fig:Figure03}c), whereas in opaque media neither the regression nor classification neural networks predict the thickness correctly.
    e) {Exemplary} projection images for bulk thicknesses or depths $d_z=8$, $d_z=16$, and $d_z=32$.
    Due to the averaging, the contrast of the waves decreases with increasing depths.
    }
  \label{fig:regression}
\end{figure}

It is possible to estimate the thickness or depth $d_z$ of transparent bulks from projections of the corresponding scroll wave dynamics using either a regression or classification neural network. 
By contrast, it is not possible to reliably predict the thickness of opaque bulks using either approach.
Fig.~\ref{fig:regression}a) shows predictions obtained with a regression neural network in transparent media, which estimates the depth accurately with floating point precision (with a certain degree of uncertainty).
Fig.~\ref{fig:regression}c) shows a confusion matrix with depth predictions obtained with a classification neural network also in transparent media, which performs better than the regression.
Out of $\sim500$ predictions per thickness, only few attempts falsely classify the thickness (off-diagonal values).
For both panels a) and c), predictions were made from two-dimensional observations as shown in Figs.~\ref{fig:Figure03}c) and \ref{fig:regression}e). 
In particular, Fig.~\ref{fig:regression}e) shows how the contrast of the waves decreases with increasing bulk thickness as more and more waves are superimposed and the signal is averaged along the $z$-axis.
The neural network presumably associates the bulk's thickness with the contrast.

Both the regression and classification neural networks are not able to predict the depth of opaque media correctly, as shown in Figs.~\ref{fig:regression}b,d). 
The classification neural network predicts random thicknesses and fails completely at correctly classifying the bulk's actual thickness.
For both panels b) and d), predictions were made from two-dimensional observations as shown in Fig. \ref{fig:Figure03}a) or b).
The data demonstrates that predicting the extend of scroll wave chaos is challenging with opacity, at least with our methodology.
While scroll wave dynamics can vary qualitatively with different bulk thicknesses, in particular with thinner bulks, as shown in Figs.~\ref{fig:Figure03}-\ref{fig:ScrollWaveChaosCrosssections}, there is a critical thickness beyond which the dynamics are dominated by the intrinsic excitable kinetics and are less influenced by the bulk's geometry and its boundaries, thus making depth predictions from surface observation challenging.

\section{Discussion}
We demonstrated that deep neural networks can be used to reconstruct three-dimensional scroll wave {dynamics from two-dimensional observations of the dynamics on the surface of excitable media.
Reconstructions succeed throughout fully opaque excitable media when the scroll wave size is not much smaller than the medium's thickness.
In that case, scroll waves can take on complex shapes with their vortex filaments being arbitrarily oriented in space, producing significant dissociation between the surface dynamics on two opposing surfaces of the medium.
However, multiple layers of scroll waves are challenging to reconstruct using encoding-decoding convolutional neural networks in opaque media, even when the dynamics are analyzed from two opposing surfaces of the medium.
Reconstructions can be performed particularly well in transparent anisotropic excitable media, in which it is possible to reconstruct also complicated scroll wave chaos far into the medium.}

{That encoding-decoding convolutional neural networks can reconstruct three-dimensional dynamics from two-dimensional observations is facilitated by their training on tens of thousands of similar examples. 
Further generalization can be achieved by diversifying the training dataset, e.g. by adding simulations with a broad range of parameters to the training data or by performing data augmentation.
The amount of information that encoding-decoding convolutional neural networks can extract from the short sequence of snapshots to perform the 2D-to-3D prediction task is remarkable.
However, it is also revealing that the dual-surface reconstruction does not provide any benefit or synergistic effects over the single-surface reconstruction.
It is as if the network performs two separate reconstructions from either side. 
This highlights fundamental limitations of convolutional encoding-decoding neural networks in this particular application.}

One interesting detail we found is that in transparent excitable media the {reconstruction outcomes are very good with anisotropy, but poor with isotropy. 
Scroll wave chaos cannot be reconstructed at all in transparent isotropic excitable media, and reconstructions of simpler scroll wave dynamics exhibit artifacts.}
These findings show that anisotropy is crucial because it implicitly encodes depth.
The neural network learns to associate the alignment of the waves with the underlying fiber alignment which varies with depth. 
Moreover, it is able to decode this encoding even when multiple waves are superimposed in the projections, see Figs.~\ref{fig:Figure03}c), \ref{fig:scrollwavechaoslayers}b) and \ref{fig:regression}e) and Supplementary Video 3.
Accordingly, the reconstructions exhibit artifacts or fail entirely in isotropic transparent excitable media as the network lacks depth information. 
Presumably, it would equally fail in anisotropic excitable media with uniform linearly transverse anisotropy.
Unfortunately, this means that this feature cannot be exploited and is neither directly applicable to ventricular fibrillation, because the ventricular muscle is opaque, nor to the Belousov-Zhabotinsky reaction, which is transparent but isotropic. 

{Nevertheless, our methodology could still be used to reconstruct intramural action potential waves including scroll waves inside cardiac tissue: 
1) Transillumination imaging \cite{Baxter2001, Bernus2007, Mitrea2009, Khait2006, Caldwell2015}, near-infrared optical mapping \cite{Hansen2018}, or other optical techniques \cite{Hillman2007, Sacconi2022}, which allow imaging of action potential waves deeper inside cardiac tissue, could be used in the transilluminated right ventricle, in the atria or in small animal hearts such as mouse or zebrafish hearts. 
The muscle fiber architecture in the projections of the transilluminated translucent tissues could enable the depth encoding.
2) Dual-surface imaging with superficial electrode mapping or fluorescent dyes lacking the penetration depth (such as Di-4-ANEPPS) could be used to reconstruct `laminar' episodes of ventricular tachycardia or atrial fibrillation. 
The wavelengths of single scroll waves or macro-reentries during ventricular arrhythmias are larger than the thickness of the right and left ventricular walls.
The atria, which exhibit epi- and endocardial dissociation during atrial fibrillation \cite{Schuessler1993, Eckstein2010, Eckstein2013, deGroot2016, Walters2020}, are presumably thin enough for dual-surface reconstructions to succeed.
However, the training data would have to account for the complex anatomy of the atria \cite{Zhao2015} as well as the particular wave dynamics. 
Whether it will be possible to create ground-truth data or to train a neural network on simulated data and subsequently apply it to experimental data needs to be determined in future research.
We found in previous work that the latter approach is in principle feasible, see Figs.~7 and 8 in \cite{Lebert2021}.
Lastly, the deep learning-based reconstructions can be performed very efficiently within milliseconds on a graphics processing unit, 
and they do not require the collection of long time-series data.}

We have recently used similar encoding-decoding convolutional neural networks for the prediction of electrical scroll wave chaos from three-dimensional mechanical deformation \cite{Christoph2020}, as well as for the prediction of phase maps and phase singularities from two-dimensional electrical spiral wave chaos \cite{Lebert2021}.
While the networks performed very well in these applications, some of the results presented in this study, particularly the results for scroll wave chaos, are more sobering.
Our study is another example of the more general notion that cardiac dynamics, and chaotic dynamics more generally, are challenging to predict \cite{Pathak2018, Pathak2018Chaos, Herzog2018, Herzog2021, Shahi2021, Shahi2022}.
It is well known that classical deep learning approaches excel at interpolating, but do not perform well at extrapolating, which is what we aimed to do in this study.
Therefore, the complete reconstruction of {complicated fine-scaled scroll wave dynamics} from surface observations in opaque excitable media will require more sophisticated {techniques than encoding-decoding convolutional neural networks}.

\section{Conclusions}
We demonstrated that it is possible to reconstruct three-dimensional scroll wave {dynamics} from two-dimensional observations using deep encoding-decoding {convolutional} neural networks.
Reconstructions succeed under two conditions: either i) the medium is transparent and anisotropic with spatially varying anisotropy or ii) the medium is opaque and the dynamics are observed on two opposing surface layers while the scroll wavelength is not much shorter than the medium's thickness.
In the future, our methodology could be used to reconstruct transmural action potential wave dynamics from epicardial or endocardial measurements.

\bibliography{references}

\section{Data Availability Statement}
The data that support the findings of this study are available from the corresponding author upon reasonable request.

\section{Funding}
This research was funded by the University of California, San Francisco, the National Institutes of Health (DP2HL168071) and the Sandler Foundation (to JC). The RTX A5000 GPUs used in this study were donated by the NVIDIA Corporation via the Academic Hardware Grant Program (to JL and JC).

\section{Author Contributions}
JL and JC conceived the research and implemented the algorithms.
JL, MM, and JC conducted the data analysis and designed the figures.
JL and JC wrote the manuscript.

\section{Conflict of Interest}
The authors declare that the research was conducted in the absence of any commercial or financial relationships that could be construed as a potential conflict of interest.

\section*{Supplementary Material}

\renewcommand{\thefigure}{S\arabic{figure}}
\setcounter{figure}{0}
\begin{figure*}[htb]
    \centering
    \includegraphics[width=0.9\textwidth]{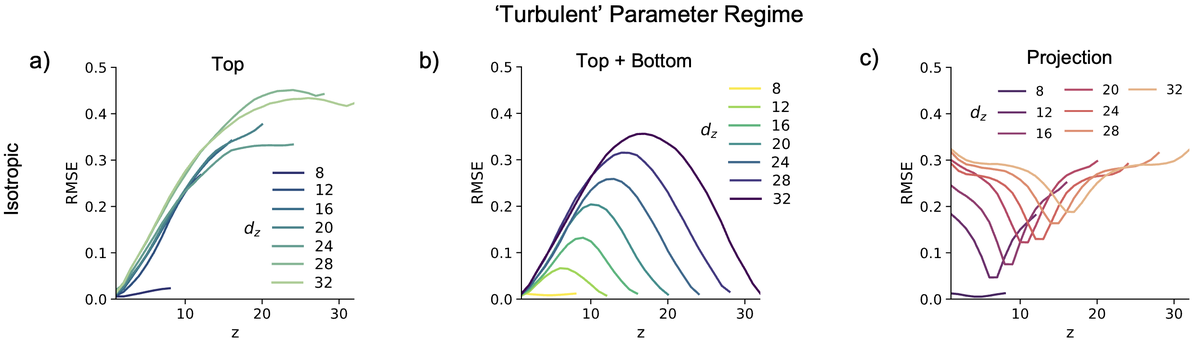}
    \caption{      
      {
      Average prediction error over bulk depth (RMSE: root mean squared error along $z$-axis) for isotropic `turbulent' simulations, see also Fig~10.
      Left: RMSE curve for top surface observations shows a steep increase in the prediction error along $z$.
      Center: Predictions in opaque anisotropic media in dual-surface mode. The error increases linearly from both sides and reaches a maximum at midwall.
      The overall error and maximum error increases with the thickness of the bulk.
      Right: RMSE curve for transparent isotropic media using $5$ projection frames. The reconstructions fail entirely, see also Fig.~8.
      For opaque single- or dual-surface observations, we see no significant difference in the prediction accuracy between isotropic and anisotropic media.
      }
    }
\end{figure*}

\begin{figure}[htb]
    \centering
    \includegraphics[width=0.48\textwidth]{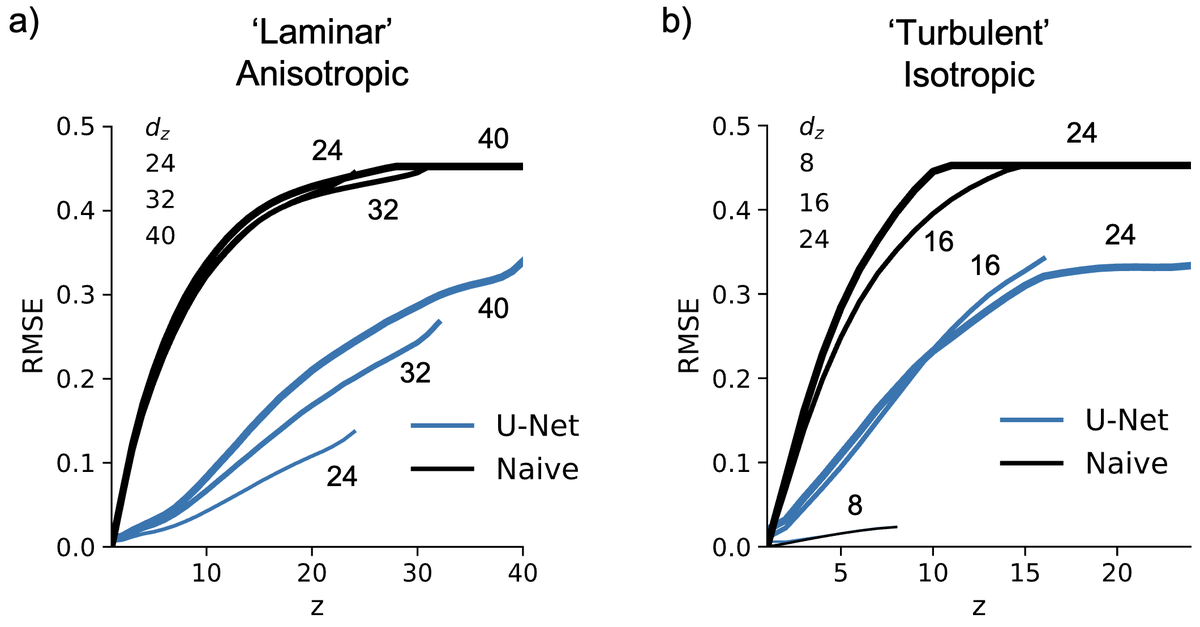}
    \caption{
{
Comparison of reconstruction error for single-surface top layer predictions along $z$-axis obtained with U-Net (blue) vs. simple extrapolation (black, Naive) by repeating the top layer as $\hat{u}(x, y, z) = u(x, y, 1)$ in all subsequent layers.
a) RMSE curves for anisotropic `laminar' parameter set simulations with $d_z=24, 32, 40$,  b) for isotropic `turbulent' simulations with $d_z=8, 16, 24$.
For the isotropic `turbulent' simulations with $d_z=8$ the dynamics are essially two-dimensional and the accuracy of the neural network and the naive estimate is almost identical.
For other $d_z$ and the `laminar' simulations the neural network predictions are significantly better than the naive estimate.
}
}
    \label{fig:NaiveEstimation}
\end{figure}

The Supplementary Videos can be found online at \href{https://gitlab.com/janlebert/subsurface-supplementary-materials}{gitlab.com/janlebert/subsurface-supplementary-materials} and at \href{https://youtube.com/@cardiacvision}{youtube.com/@cardiacvision} under `Playlists' $\rightarrow$ `Paper: Reconstruction of Three-dimensional Scroll Waves in Excitable Media from Two-Dimensional Observations using Deep Neural Networks'.
$\\$
$\\$
{
\textbf{Supplementary Video 1:} Deep-learning based reconstruction of scroll wave dynamics in a bulk-shaped excitable medium from two-dimensional observations of the dynamics on the bulk's surface, see also Fig.~\ref{fig:ScrollWave}. The observations are either 1) superficial observations of a single surface ("Top Prediction") in an opaque excitable medium, or 2) two opposing surfaces ("Top + Bottom Prediction") in an opaque excitable medium, or 3) a projection of the fully three-dimensional dynamics along the $z$-axis ("Projection Prediction") in a transparent excitable medium. In all 3 cases, the excitable medium is anisotropic and the thickness of the bulk is $d_z=24$ voxels.}
$\\$
$\\$
{
\textbf{Supplementary Video 2:} Deep-learning based reconstruction of scroll wave dynamics in a thicker bulk-shaped excitable medium from two-dimensional observations of the dynamics on the bulk's surface, see also Fig.~\ref{fig:LargeScrollWave}. The observations are either 1) superficial observations of a single surface ("Top Prediction") in an opaque excitable medium, or 2) two opposing surfaces ("Top + Bottom Prediction") in an opaque excitable medium, or 3) a projection of the fully three-dimensional dynamics along the $z$-axis ("Projection Prediction") in a transparent excitable medium. In all 3 cases, the excitable medium is anisotropic and the thickness of the bulk is $d_z=40$ voxels. In thicker bulks, it is increasingly difficult to reconstruct the dynamics in deeper layers and throughout the entire bulk when analyzing only one of the surfaces.}
$\\$
$\\$
\textbf{Supplementary Video 3:} {Deep-learning based reconstruction of three-dimensional `turbulent' scroll wave dynamics from two-dimensional projections of the dynamics in transparent anisotropic excitable medium. The three-dimensional reconstructions are computed from $5$ two-dimensional snapshots.}
The projection frame for each time point is shown on the left, the neural network prediction on the right. The true full three-dimension excitation is shown on the left in the second half of the video. The thickness of the bulk is $d_z=24$ voxels.
$\\$
$\\$
\textbf{Supplementary Video 4:} {Comparison of deep-learning based three-dimensional reconstructions of `turbulent' scroll wave dynamics from two-dimensional observations, see also Figs.~\ref{fig:ScrollWaveChaosCrosssections} and \ref{fig:scrollwavechaoslayers}. Top left: Ground-truth, top right: projection in transparent medium, bottom left: top layer in opaque medium, bottom right: top and bottom layer in opaque medium.} The medium is anisotropic and the thickness of the bulk is $d_z=24$ voxels. In the second half of the video only the bottom $12$ layers are visualized.
$\\$
$\\$
{
\textbf{Supplementary Video 5:} Deep-learning based reconstruction with noise, see also Fig.~\ref{fig:Noise}. The reconstruction of three-dimensional scroll wave dynamics from two-dimensional projections succeeds with mild ($\sigma=0.05$) and strong noise ($\sigma=0.2$) in transparent excitable media. Left: Original projection without noise. Center: projections with noise. Right: Reconstruction.}

\end{document}